\documentclass[12pt]{article}
\pdfoutput=1
\usepackage{amssymb,amsmath}
\usepackage{epsfig}
\usepackage{epstopdf}
\usepackage{latexsym}
\usepackage{graphicx}
\usepackage{booktabs}
\usepackage{bbm}
\usepackage{enumitem}
\usepackage[numbers,compress]{natbib}
\usepackage{float}
\numberwithin{equation}{section}

\usepackage[T1]{fontenc}

\usepackage[margin=20pt,small]{caption}
\usepackage{subcaption}

\usepackage[toc]{appendix}

\usepackage{tikz}
\usetikzlibrary{decorations.pathreplacing}

\usepackage{array}
\usepackage[all]{xy}

\DeclareCaptionType{equ}[Diagram][]

\usepackage{color}
\usepackage{datetime}
\usepackage[
      colorlinks=false,
      linkcolor=darkblue,  
      urlcolor=blue,    
      filecolor=blue,     
      citecolor=red,
linktocpage=true,
      pdfstartview=FitV,
      bookmarksopen=true,
	  hidelinks
      ]{hyperref}

\ifpdf
\fi

\newcommand*{\boxedcolor}{red}
\makeatletter
\renewcommand{\boxed}[1]{\textcolor{\boxedcolor}{%
  \fbox{\normalcolor\m@th$\displaystyle#1$}}}
\makeatother

\definecolor{cardinal}{rgb}{0.6,0,0}
\definecolor{darkgreen}{rgb}{0,0.5,0}
\definecolor{golden}{rgb}{0.92, 0.7, 0}
\definecolor{midnight}{rgb}{0, 0, 0.5}
\definecolor{darkblue}{rgb}{0.2, 0, 0.8}

\newcommand{\dd}{\mathrm{d}}

\newcommand{\e}{\mathrm{e}}
\newcommand{\rmi}{\mathrm{i}}
\newcommand{\rme}{\mathrm{e}}
\newcommand{\rmd}{\mathrm{d}}
\newcommand{\w}{\wedge}

\newcommand{\Tr}{\text{Tr}~}

\newcommand{\SU}{\mathop{\rm SU}}
\newcommand{\SO}{\mathop{\rm SO}}

\newcommand{\SL}{\mathop{\rm SL}}
\newcommand{\U}{\mathop{\rm {}U}}
\newcommand{\USp}{\mathop{\rm {}USp}}

\def\cM{{\cal M}}
\def\cN{{\cal N}}

\def\cV{{\cal V}}

\def\Tr{{\rm Tr}\,}

\def\SL{{\rm SL}}

\def\SO{{\rm SO}}

\def\SU{{\rm SU}}

\newcommand{\gM}{{\cal M}}
\newcommand{\obf}[1]{\overline{\mathbf{#1}}}
\newcommand{\gL}{\mathcal{L}}
\newcommand{\uM}{\underline{M}}
\newcommand{\uN}{\underline{N}}
\newcommand{\uP}{\underline{P}}

\newcommand{\uA}{\underline{A}}
\newcommand{\uB}{\underline{B}}
\newcommand{\uC}{\underline{C}}
\newcommand{\uD}{\underline{D}}
\newcommand{\uE}{\underline{E}}
\newcommand{\uF}{\underline{F}}
\newcommand{\UI}{\left(U^{-1}\right)}
\newcommand{\gA}{{\cal A}}
\newcommand{\gB}{{\cal B}}
\newcommand{\cU}{{\cal U}}

\newcommand\mc[1]{{\mathcal{#1}}}

\newcommand\vphi{\varphi}

\newcommand\wt[1]{{\widetilde{#1}}}

\def\shrug{\texttt{\raisebox{0.75em}{\char`\_}\char`\\\char`\_\kern-0.5ex(\kern-0.25ex\raisebox{0.25ex}{\rotatebox{45}{\raisebox{-.75ex}"\kern-1.5ex\rotatebox{-90})}}\kern-0.5ex)\kern-0.5ex\char`\_/\raisebox{0.75em}{\char`\_}}}

\newcommand{\be}{\begin{equation}}
\newcommand{\ee}{\end{equation}}
\newcommand{\bea}{\begin{eqnarray}}
\newcommand{\eea}{\end{eqnarray}}

\newcommand{\mbf}{\mathbf}

\begin{document}  

\begin{titlepage}
\begin{flushright} \small
HU-EP-20/41
\end{flushright}

\medskip
\begin{center} 
{\Large \bf  Kaluza-Klein Spectroscopy for the Leigh-Strassler SCFT}

\bigskip
\bigskip
\bigskip
\bigskip

{\bf Nikolay Bobev,${}^{\rm KK}$ Emanuel Malek,${}^{\rm L}$ Brandon Robinson,${}^{\rm KK}$ Henning Samtleben,${}^{\rm S}$ and Jesse van Muiden${}^{\rm KK}$  \\ }
\bigskip
\bigskip
\bigskip
\bigskip
${}^{\rm KK}$Instituut voor Theoretische Fysica, KU Leuven \\
Celestijnenlaan 200D, B-3001 Leuven, Belgium
\vskip 8mm
${}^{\rm L}$Institut f\"ur Physik, Humboldt-Universit\"at zu Berlin\\
 IRIS Geb\"aude, Zum Gro\ss en Windkanal 2, 12489 Berlin, Germany
 \vskip 8mm
 ${}^{\rm S}$Univ Lyon, Ens de Lyon, Univ Claude Bernard, CNRS,\\
 Laboratoire de Physique, F-69342 Lyon, France\\
 
\bigskip
\texttt{nikolay.bobev,~brandon.robinson,~jesse.vanmuiden~@kuleuven.be, emanuel.malek@physik.hu-berlin.de, henning.samtleben@ens-lyon.fr} \\
\end{center}

\bigskip
\bigskip

\begin{abstract} 

\noindent We apply recently developed tools from exceptional field theory to calculate the full Kaluza-Klein spectrum of the AdS$_5$ Pilch-Warner solution of type IIB supergravity. Through the AdS/CFT correspondence this yields detailed information about the spectrum of protected and unprotected operators of the four-dimensional $\cN=1$ Leigh-Strassler SCFT, in the planar limit. We also calculate explicitly the superconformal index of the SCFT in this limit and show that it agrees precisely with the spectrum of protected operators in the supergravity calculation. 
  
\end{abstract}

\noindent

\end{titlepage}


\setcounter{tocdepth}{2}
\tableofcontents


\section{Introduction}

The spectrum of conformal dimensions of local operators is the basic data in any conformal field theory. Unfortunately, there is a shortage of tools for the exact calculation of the conformal dimensions for interacting CFTs in more than two dimensions. For theories with enough supersymmetry the dimensions of certain BPS operators can be computed exactly by exploiting the fact that they belong to special short, or protected, multiplets of the superconformal algebra. Finding the spectrum of non-BPS operators however is usually out of reach. The goal of this paper is to calculate the conformal dimensions of infinitely many protected and unprotected local operators in a 4d $\mathcal{N}=1$ theory known as the Leigh-Strassler (LS) SCFT \cite{Leigh:1995ep}. To render the calculation manageable we focus on the LS theory with gauge group $\SU(N)$ and study it in the large $N$ limit where it can be described by a holographically dual AdS$_5$ solution of type IIB supergravity.

The LS SCFT can be obtained as an IR fixed point arising from the $\mathcal{N}=4$ SYM theory via an RG flow triggered by a relevant deformation. More specifically one can write the $\mathcal{N}=4$ SYM theory in an $\mathcal{N}=1$ formulation with the following superpotential for the three chiral superfields
\begin{equation}\label{eq:Wintro}
W = {\rm Tr}\, \Phi_1 [ \Phi_2,\Phi_3 ] + \frac{m}{2}\Phi_1^2\,.
\end{equation}
For $m=0$ one finds the $\mathcal{N}=4$ SCFT. The RG flow triggered by the relevant coupling $m\neq 0$ ends in a non-trivial 4d $\mathcal{N}=1$ SCFT with an effective quartic superpotential for $\Phi_2$ and $\Phi_3$ which can be obtained by integrating out $\Phi_1$ \cite{Leigh:1995ep}. The LS SCFT has an $\SU(2)_F$ flavor symmetry and a $\U(1)_R$ superconformal R-symmetry. It is a part of a three-dimensional conformal manifold, which is believed to be compact and thus does not have a weakly coupled locus. This should be contrasted with the $\mathcal{N}=4$ SYM theory which has a free limit and one can gain some calculational control of the operator spectrum by employing perturbation theory. The fact that the LS theory is intrinsically strongly interacting complicates the calculation of the spectrum of operator dimensions. 

The AdS/CFT correspondence has proven to be an invaluable tool in addressing such strongly coupled CFT problems and indeed it can be brought to bear for the LS SCFT. The holographic description of the RG flow connecting $\mathcal{N}=4$ SYM and the $\mathcal{N}=1$ LS fixed point is given by a domain wall solution that interpolates between two supersymmetric AdS$_5$ vacua of the five-dimensional, $\SO(6)$ gauged, $\mathcal{N}=8$ supergravity \cite{Freedman:1999gp}. The AdS$_5$ vacuum dual to the UV $\mathcal{N}=4$ fixed point is the maximally supersymmetric $\SO(6)$ invariant vacuum of the 5d theory \cite{Gunaydin:1984qu,Gunaydin:1985cu,Pernici:1985ju}. The IR AdS$_5$ vacuum preserves $1/4$ of the maximal supersymmetry as well as an $\SU(2)_F\times \U(1)_R$ subgroup of $\SO(6)$ and was found in \cite{Khavaev:1998fb} as a non-trivial critical point of the scalar potential of the 5d gauged supergravity. Importantly, the full holographic RG flow and the IR AdS$_5$ vacuum can be explicitly uplifted to solutions of type IIB supergravity \cite{Pilch:2000ej,Pilch:2000fu}. The existence of this uplift to ten dimensions is now understood to be a consequence of the fact that the 5d $\SO(6)$ $\mathcal{N}=8$ supergravity can be obtained as a consistent truncation of type IIB supergravity on $S^5$ \cite{Lee:2014mla,Baguet:2015sma}. One can hope to use the explicit 10d AdS$_5$ Pilch-Warner (PW) solution in \cite{Pilch:2000ej} to study the spectrum of perturbative KK excitations as was done for the maximally supersymmetric AdS$_5\times S^5$ solution in~\cite{Kim:1985ez,Gunaydin:1984fk,vanNieuwenhuizen:2012zk}. The masses of the KK modes can then mapped to the conformal dimensions of  local operators in the dual SCFT. This however proves to be a hard problem. The reason is that the solution of \cite{Pilch:2000ej} is not of Freund-Rubin type, i.e. the metric on~$S^5$ is not the round one and there are non-trivial fluxes for the R-R and NS-NS 2-forms on the~$S^5$. This in turn implies that the group theory techniques used in \cite{Kim:1985ez,Gunaydin:1984fk,vanNieuwenhuizen:2012zk} cannot be directly applied to the PW supergravity solution.

A  new tool was recently  developed to attack precisely such KK spectroscopy problems~\cite{Malek:2019eaz,Malek:2020yue}. An important stepping stone that facilitated this development is the formulation of exceptional field theory (ExFT) as a duality covariant formulation of 10d or 11d supergravity \cite{Hohm:2013pua}. In the context of type IIB supergravity, the ExFT formalism allows for a formulation of 10d supergravity in an $\rm E_{6(6)}$ covariant way which, is tailor made for reductions to 5d \cite{Hohm:2014qga}. An important application of this formalism leads to a proof that the 5d $\mathcal{N}=8$ $\SO(6)$ gauged supergravity theory of \cite{Gunaydin:1984qu,Gunaydin:1985cu,Pernici:1985ju} is a consistent truncation of type IIB supergravity on $S^5$ \cite{Baguet:2015sma}. In addition to proving this important fact the results in \cite{Baguet:2015sma} provide explicit equations that relate the 5d and 10d supergravity fields that can be used to uplift any solution of the equations of motion in the 5d theory to a type IIB background.\footnote{See \cite{Bobev:2018hbq,Petrini:2018pjk,Bobev:2018eer,Bobev:2019wnf,Bobev:2020fon} for several applications of the uplift formulae of \cite{Baguet:2015sma} in a holographic context.} It was shown in \cite{Malek:2019eaz,Malek:2020yue} that these consistent truncation results can be extended to apply not only to a given 5d solution and its 10d uplift but also to the full spectrum of quadratic fluctuations around a given supergravity solution. Using the results in \cite{Malek:2019eaz,Malek:2020yue} one can therefore find the KK spectrum of any type IIB supergravity background which arises as an uplift of a solution in the 5d $\mathcal{N}=8$ supergravity.\footnote{See  \cite{Malek:2020mlk,Guarino:2020flh,Cesaro:2020soq} for recent applications of the results in \cite{Malek:2019eaz,Malek:2020yue} for calculating the KK spectrum of AdS solutions in 4d maximal supergravity theories, and \cite{Eloy:2020uix} for application to AdS$_3$ vacua.}  The AdS$_5$ supergravity dual to the LS SCFT is precisely such a solution, and thus we can readily apply the methods in \cite{Malek:2019eaz,Malek:2020yue}. Indeed, a main result of this work is to show how to apply the formalism in \cite{Malek:2019eaz,Malek:2020yue} to calculate the KK spectrum of the PW solution \cite{Pilch:2000ej} and organize it in 4d $\mathcal{N}=1$ superconformal multiplets. We find that the results can be conveniently packaged in terms of generating functions that encode all the
information about the superconformal representation theory accompanied by explicit analytic formulae
for all operator dimensions. Importantly, we find that the KK spectrum comprises of modes dual to both protected and unprotected operators in the LS SCFT. This is in contrast to the KK spectrum of the maximally supersymmetric AdS$_5\times S^5$ solutions, where all KK modes are dual to protected operators, and is akin to the KK spectrum of the AdS$_5\times T^{1,1}$ solution dual to the $\mathcal{N}=1$ KW SCFT \cite{Romans:1984an,Klebanov:1998hh,Gubser:1998vd,Ceresole:1999ht,Ceresole:1999zs}.

Some of these ExFT results for the KK spectrum can be derived independently by a more direct calculation using the 10d PW AdS$_5$ solution \cite{Pilch:2000ej}. One can show, similarly to \cite{Klebanov:2009kp}, that the spectrum of spin-2 excitations around this 10d solution is equivalent to that of a minimally coupled scalar field. The corresponding eigenfunctions and eigenvalues for the wave equation in 10d can be found explicitly. The results for the spin-2 KK spectrum obtained in this way are in perfect agreement with the ExFT results.

The spectrum of protected operators in 4d $\mathcal{N}=1$ SCFTs can be found by calculating the superconformal index \cite{Romelsberger:2005eg,Kinney:2005ej}. This index is relatively easy to compute when the theory of interest has a weakly coupled description, like $\mathcal{N}=4$ SYM. This is due to the fact that the index is invariant under the continuous deformation by a marginal coupling. As discussed in \cite{Gadde:2010en}, this argument can also be applied for theories related by an RG flow to a theory admitting a weakly coupled limit. We can utilize this approach for the LS theory and compute its superconformal index. In particular, we can use the results for the large $N$ limit of the index in \cite{Gadde:2010en} to arrive at a compact closed form expression for the so-called single-trace index which captures all contributions from protected single-trace local operators dual to KK supergravity modes. We show how the simple expression for the single-trace index can be expanded as a collection of 8 infinite series each of which captures the contribution from KK towers of semi-short superconformal multiplets. The agreement between this superconformal index calculation and the KK spectroscopy results amounts to an explicit detailed test of the holographic duality between the PW AdS$_5$ solution and the LS SCFT.

In the next section we summarize the 5d AdS$_5$ supergravity solution dual to the LS SCFT and show how to compute the KK spectrum around it, using the technique developed in \cite{Malek:2019eaz,Malek:2020yue}. In Section~\ref{sec:IIB} we present the type IIB supergravity uplift of this 5d solution, compute the KK spectrum of spin-2 excitations by a direct calculation, and show that the results agree with those using the ExFT technique. The superconformal index of the LS theory is computed in Section~\ref{sec:index} where we also show explicitly how it reproduces the spectrum of protected operators derived in supergravity. We conclude in Section~\ref{sec:discussion} with a discussion of some open questions. The two appendices contain a summary of some results in 4d $\mathcal{N}=1$ superconformal representation theory, as well as explicit expressions for the first 4 levels of the KK spectrum of the PW solution.

\section{5d supergravity, ExFT, and KK spectroscopy}
\label{sec:5dKK}

We start our exploration by describing the 5d supergravity solution dual to the 4d $\mathcal{N}=1$ LS fixed point. The AdS$_5$ solution dual to the conformal vacuum of the LS theory was found in \cite{Khavaev:1998fb} and amounts to a non-trivial critical point of the scalar potential of the 5d $\mathcal{N}=8$ $\SO(6)$ gauged supergravity constructed in \cite{Gunaydin:1984qu,Gunaydin:1985cu,Pernici:1985ju}. The holographic description of the RG flow from $\mathcal{N}=4$ SYM to the $\mathcal{N}=1$ fixed point is in terms of a domain wall solution interpolating between two AdS$_5$ supersymmetric vacua of the supergravity theory and was discussed in detail in \cite{Freedman:1999gp}. 

To understand better this RG flow it is useful to describe its symmetries. The $\mathcal{N}=4$ SYM theory can be formulated in $\mathcal{N}=1$ language with superpotential $W = {\rm Tr}\, \Phi_1 [ \Phi_2 , \Phi_3 ] $ as in \eqref{eq:Wintro}. In this formulation only an $\SU(3) \times \U(1)_R^{\rm UV}$ subgroup of the $\SO(6)$ R-symmetry is manifest. Here $\U(1)_R^{\rm UV}$ is the superconformal R-symmetry in the UV and is given by the following linear combination of the Cartan generators of $\SO(6)$
\begin{equation}\label{eq:TRUV}
T_R^{\rm UV} = \frac{2}{3}(T_1+T_2+T_3)\,,
\end{equation}
where $T_{1,2,3}$ are the generators of the 3 $\SO(2)$ block-diagonal blocks in $\SO(6)$. By adding the superpotential mass term $\Delta W = \frac{m}{2}\Phi_1^2$ in \eqref{eq:Wintro} the global symmetry is broken to $\SU(2)_F \times \U(1)_R^{\rm IR}$, where $\U(1)_R^{\rm IR}$ is the IR superconformal R-symmetry given explicitly by the linear combination
\begin{equation}\label{eq:TRIR}
T_R^{\rm IR} = \frac{1}{2}(T_1+T_2+2T_3)\,.
\end{equation}
The $\SU(2)_F$ flavor symmetry is the first factor in the $\SU(2)_1\times \SU(2)_2\times \U(1)_{56}$ subgroup of $\SO(6)$, where $\U(1)_{56}$ is generated by $T_3$.

As discussed in \cite{Intriligator:1998ig}, correlation functions of local operators in the planar limit of the $\mathcal{N}=4$ SYM enjoy a ``bonus'' $\U(1)_Y$ symmetry. This $\U(1)_Y$ is the compact subgroup of the $\SL(2,\mathbb{R})$ symmetry of the 5d supergravity theory which in turn is the large $N$ manifestation of the $\SL(2,\mathbb{Z})$ S-duality group of the $\mathcal{N}=4$ SYM theory. This ``bonus'' $\U(1)_Y$ symmetry plays an important role in our discussion below.

One can use these symmetries to show that the holographic RG flow of interest can be constructed by using only 2 of the 42 scalars, which we denote by $(\alpha,\chi)$, in the 5d $\mathcal{N}=8$ gauged supergravity theory. The scalar $\alpha$ is dual to a scalar bilinear operator in the $\mathbf{20'}$ representation of $\SO(6)$, while $\chi$ is dual to a linear combination of fermionic bilinear operators in the $\mathbf{10}\oplus\overline{\mathbf{10}}$. These two operators are precisely the ones that are being sourced by the superpotential mass term in \eqref{eq:Wintro}.

To describe the supergravity domain wall of interest we focus on the scalar sector of the 5d $\mathcal{N}=8$ supergravity theory. It consists of 42 scalars spanning the coset
\begin{equation}\label{eq:E6coset}
\frac{\rm E_{6(6)}}{\USp(8)}\,.
\end{equation}
The infinitesimal generators of the coset can be parametrized as \cite{Gunaydin:1985cu}
\begin{equation}\label{eq:genE6}
\mathfrak g= \left( 
\begin{array}{cc}
-4 \Lambda^{[i}_{\phantom{[i}[k}\delta^{j]}_{l]} & \sqrt{2} \Sigma_{ijk\alpha}  \\ 
 \sqrt{2} \Sigma^{ijk\alpha} & \Lambda^i_{\phantom{i}j}\delta^{\alpha}_{\beta} + \Lambda^\alpha_{\phantom{\alpha}\beta}\delta^{i}_{j}
\end{array}  \right)\,,
\end{equation}
where $\Lambda^i_{\phantom{i}j}$ is symmetric traceless matrix that generates $\SL(6,\mathbb{R})$, while $\Lambda^{\alpha}_{\phantom{\alpha}\beta}$ is a symmetric traceless matrix that generates $\SL(2,\mathbb{R})$. The off-diagonal blocks in \eqref{eq:genE6} are fully anti-symmetric in $(i,j,k)$, and self-dual, i.e. they obey the relation
\begin{equation}
\Sigma_{ijk\alpha} = \frac16 \varepsilon_{\alpha\beta}\,\varepsilon_{ijklmn}\,\Sigma^{lmn\beta}\,.
\end{equation}
For the two-scalar truncation of interest these tensors simplify significantly, see Appendix A in \cite{Freedman:1999gp}. For the diagonal blocks in \eqref{eq:genE6} we have
\begin{equation}
\begin{aligned}
\Lambda^i_{\phantom{i}j} =&\, \text{diag}\left[ -\alpha,-\alpha,-\alpha,-\alpha,2\alpha, 2\alpha \right]\,, \qquad \Lambda^\alpha_{\phantom{\alpha}\beta} =&0\,.
\end{aligned}
\end{equation}
The off-diagonal blocks are given by
\begin{equation}
\Sigma^{ij51} = \Sigma^{ij62} = \frac{\chi }{{2}} 
\left[\begin{array}{cc}
 \rmi \sigma_2& 0  \\ 
0 & \rmi \sigma_2
\end{array} \right]^{ij}
\,,\qquad  \Sigma^{ij61} = -\Sigma^{ij52} = -\frac{ \chi }{{2}} \left[\begin{array}{cc}
 0& \sigma_3  \\ 
\sigma_3 & 0
\end{array} \right]^{ij}\,,
\end{equation}
where $\sigma_a$ are the Pauli matrices. The metric on the scalar coset is given by
\begin{equation} \label{eq:5dGenMetric}
M = \mathcal V \cdot \mathcal V^T\,,\quad \text{where} \quad  \mathcal V = \exp \left[ \mathfrak g \right]\,,
\end{equation}
and can be used to write down the Lagrangian of the scalar sector of the 5d $\mathcal{N}=8$ supergravity theory as
\begin{equation}
\begin{aligned}
\mathcal L =& \sqrt{\left| g_{5} \right|}\left( R_5 -\mathcal K + \mathcal P \right)  \,.
\end{aligned}
\end{equation}
Here we have defined the scalar kinetic term and potential as\footnote{We use underlined indices on the coset metric to be consistent with the notations used in \cite{Malek:2020yue}, and in Section~\ref{subsec:ExFT} below.}
\begin{equation}\label{Eq: 5D kinetic and potential terms}
\begin{aligned}
\mathcal K =& \, \frac{1}{24} \partial_\mu M_{\underline{M}\underline{N}}\partial^\mu M^{\underline{M}\underline{N}} = -12 \left(\alpha'\right)^2 - 2\left(\chi'\right)^2 \,,\\
\mathcal P =&\, \frac{\mathtt g^2 \cosh^2 \chi}{2\rme^{4\alpha}}\left( \rme^{12 \alpha} \sinh^2 \chi - 4 \rme^{6\alpha} + \cosh 2\chi -3 \right)\,.
\end{aligned}
\end{equation}
%
The supergravity domain wall realizing the LS RG flow has the five-dimensional metric\footnote{Note that, unlike \cite{Gunaydin:1985cu}, we work in mostly plus conventions for the metric signature.}
\begin{equation}
\rmd s_{\text{AdS}_5}^2 = \rmd r^2 + \rme^{2A} \eta_{\mu\nu}\rmd x^\mu \rmd x^\mu\,.
\end{equation}
The scalar field and the metric function $A$ depend only of the radial coordinate $r$ and obey the following system of BPS equations 
\begin{equation}
\alpha' = \frac{\mathtt g}{12}\partial_\alpha \mathcal W\,, \qquad \chi' = \frac{\mathtt g}{2} \partial_\chi \mathcal W\,, \qquad  A' = - \frac{\mathtt g}{3} \mathcal W\,.
\end{equation}
Here we have defined the superpotential function $\mathcal W$ which reads\footnote{The superpotential can be obtained as an eigenvalue of the $W_{ab}$ tensor in \cite{Gunaydin:1985cu}.}
\begin{equation}
\mathcal W = \frac{\rme^{4\alpha}}{4}\left(\cosh2\chi-3\right) - \rme^{-2\alpha} \cosh^2 \chi\,.
\end{equation}
The potential is related to the superpotential as
\begin{equation}
\mathcal P = \mathtt g^2\left[\tfrac{1}{12} (\partial_\alpha \mathcal W)^2 + \tfrac{1}{2} (\partial_\chi \mathcal W)^2 - \tfrac{4}{3} \mathcal W^2\right]\,,
\end{equation}
and the supersymmetric AdS$_5$ solutions are critical points of the superpotential, $\partial_\alpha \mathcal W=\partial_\chi \mathcal W=0$, which obey
\begin{equation}
\mathcal{P} = -\frac{4\mathtt g^2}{3}  \mathcal W^2\,.
\end{equation}
The potential in \eqref{Eq: 5D kinetic and potential terms} has two supersymmetric critical points\footnote{As discussed in \cite{Krishnan:2020sfg,Bobev:2020ttg} these are the only supersymmetric AdS$_5$ solutions in the full 5d $\mathcal{N}=8$ gauged supergravity theory.} \cite{Khavaev:1998fb}. The first one is the maximally supersymmetric $\SO(6)$ invariant critical point where
\begin{equation}
\alpha = 0\,, \qquad  \chi = 0\,, \qquad \text{and} \qquad \mathcal P_{\text{UV}} = -3 \mathtt g^2\,.
\end{equation}
This is the AdS$_5$ solution dual the UV $\mathcal{N}=4$ SYM theory. The second critical point preserves $1/4$ of the maximal supersymmetry and is given by
\begin{equation}  \label{eq:PilchWarner}
\rme^{6\alpha} = 2\,,  \qquad \rme^{2\chi} = 3\,, \qquad \text{and} \qquad \mathcal P_{\text{IR}} = -\frac{8 \cdot 2^{1/3} }{3}\mathtt g^2\,.
\end{equation}
This is the AdS$_5$ dual to the LS $\mathcal{N}=1$ SCFT.

The AdS length scale is determined by the value, $\mathcal P_*$, of the potential at the critical point through the relation
\begin{equation}
\quad L^2 = -\frac{12}{\mathcal P_*}\,.
\end{equation}
For the two critical points of interest we find the explicit expressions.
\begin{equation}\label{eq:LUVIR}
L_{\text{UV}} =  \frac{2}{\mathtt g}\,, \quad \text{and} \quad   L_{\text{IR}} = \frac{3}{2^{2/3} \mathtt g}\,.
\end{equation}
We note in passing that the ratio of the conformal anomaly coefficients (in the planar limit) of the UV and IR SCFTs is given by
\begin{equation}
\frac{c_{\rm IR}}{c_{\rm UV}} = \frac{L_{\text{IR}} ^3}{L_{\text{UV}} ^3} = \frac{27}{32}\,.
\end{equation}
This is the field theory result derived in \cite{Freedman:1999gp,Tachikawa:2009tt}.

The two supersymmetric AdS$_5$ solutions can be uplifted to supersymmetric backgrounds of type IIB supergravity. The UV AdS$_5$ vacuum uplifts to the well-known AdS$_5\times S^5$ solution while the IR AdS$_5$ vacuum gives rise to the Pilch-Warner solution which we present explicitly in Section~\ref{subsec:PWsoln} below. The 5d holographic RG flow solution that  interpolates between the two supersymmetric AdS$_5$ vacua above was constructed numerically in \cite{Freedman:1999gp} and, as shown in \cite{Pilch:2000fu}, can be uplifted to type IIB supergravity.

\subsection{ExFT spectroscopy}
\label{subsec:ExFT}

The AdS$_5$ solution in \eqref{eq:PilchWarner} can be uplifted to the Pilch-Warner type IIB solution \cite{Pilch:2000ej}. This implies that we can use the ExFT methods of \cite{Malek:2019eaz,Malek:2020yue} to compute the spectrum of Kaluza-Klein fluctuations around this 10d AdS$_5$ solution. Here we summarize the results of \cite{Malek:2019eaz,Malek:2020yue}, beginning with a brief review of ExFT and its description of consistent truncations.

ExFT provides an $\rm E_{6(6)}$-covariant reformulation of 10-/11-dimensional supergravity that unifies its metric and flux degrees of freedom. The bosonic sector of ExFT consists of a 5d metric, $g_{\mu\nu}$, a generalized metric, $\gM_{MN}$, parameterizing the coset space ${\rm E_{6(6)}}/\USp(8)$, a vector $\gA_\mu{}^M$ transforming in the $\mathbf{27}$ of $\rm E_{6(6)}$ and a 2-form $\gB_{\mu\nu\,M}$, transforming in the $\obf{27}$ of $\rm E_{6(6)}$. Here the indices range as $\mu = 0, \ldots, 4$ and $M = 1, \ldots, 27$.

Consistent truncations of 10-/11-dimensional supergravity to a 5d gauged supergravity with maximal supersymmetry are elegantly captured in ExFT in terms of a generalized Scherk-Schwarz Ansatz:
\begin{equation}
	\begin{split} \label{eq:gSS}
		\gM_{MN} &= U_M{}^{\uM}\, U_N{}^{\uN}\, M_{\uM\uN}(x) \,, \\
		\gA_{\mu}{}^M &= \rho^{-1}\, \UI_{\uM}{}^M A_{\mu}{}^{\uM}(x) \,, \\
		\gB_{\mu\nu\,M} &= \rho^{-2}\, U_M{}^{\uM}\, B_{\mu\nu\,\uM}(x) \,, \\
		g_{\mu\nu} &= \rho^{-2}\, \mathring{g}_{\mu\nu}(x) \,,
	\end{split}
\end{equation}
where $U_M{}^{\uM} \in {\rm E_{6(6)}}/\USp(8)$ is called the twist matrix and $\rho \in \mathbb{R}^+$ is a scaling function, both of which depend only on the internal manifold and have to satisfy the differential condition
\begin{equation} \label{eq:GenPar}
	\gL_{\cU_{\uM}} \cU_{\uN}{}^M = X_{\uM\uN}{}^{\uP}\, \cU_{\uP} \,,
\end{equation}
in terms of the generalized Lie derivative, $\gL$, whose precise form we will not need here, and
\begin{equation}
	\cU_{\uM}{}^M = \rho^{-1} \UI_{\uM}{}^M \,.
\end{equation}
The fields $M_{\uM\uN}(x) \in {\rm E_{6(6)}}/\USp(8)$, $A_\mu{}^{\uM}(x)$, $B_{\mu\nu\,\uM}(x)$ and $\mathring{g}_{\mu\nu}(x)$ are those of the 5-dimensional gauged supergravity with the embedding tensor given by $X_{\uM\uN}{}^{\uP}$.\footnote{The explicit realization of this tensor for the $\SO(6)$ gauged supergravity we are studying here is given in \eqref{eq:S5X}.}

Here we are interested in the AdS$_5$ vacuum in \eqref{eq:PilchWarner}, whose only non-trivial 5d fields are $\mathring{g}_{\mu\nu}(x) = g_{{\rm AdS}_5}$ and $M_{\uM\uN} = {\cal V}_{\uM}{}^{\uA}\, {\cal V}_{\uN}{}^{\uB}\, \delta_{\uA\uB}$, with $\uA, \uB = 1, \ldots, 27$, and ${\cal V}$ as given in \eqref{eq:5dGenMetric}. As shown in \cite{Malek:2019eaz,Malek:2020yue}, a general KK fluctuation around such a vacuum can be written as
\begin{equation}
	\begin{split} \label{eq:FluctAnsatz}
		\gM_{MN} &= U_M{}^{\uA}\, U_N{}^{\uB} \left( \delta_{\uA\uB} + {\cal P}_{\uA\uB,I} \sum_\Sigma {\cal Y}^\Sigma\, j_{I,\Sigma}(x) \right) \,, \\
		\gA_{\mu}{}^M &= \rho^{-1}\, \UI_{\uA}{}^M \sum_\Sigma {\cal Y}^\Sigma A_{\mu}{}^{\uA,\Sigma}(x) \,, \\
		\gB_{\mu\nu\,M} &= \rho^{-2}\, U_M{}^{\uA} \sum_\Sigma {\cal Y}^\Sigma B_{\mu\nu\,\uA,\Sigma}(x) \,, \\
		g_{\mu\nu} &= \rho^{-2} \left( \mathring{g}_{\mu\nu}(x) + \sum_\Sigma {\cal Y}^\Sigma\, h_{\mu\nu,\Sigma}(x) \right) \,.
	\end{split}
\end{equation}
Let us explain the new objects appearing in \eqref{eq:FluctAnsatz}. Firstly, we have defined new twist matrices $U_M{}^{\uA}$ dressed with the 5d scalar matrix associated to the vacuum as
\begin{equation}
	U_M{}^{\uA} = U_M{}^{\uM} {\cal V}_{\uM}{}^{\uA} \,.
\end{equation}
Secondly, we have introduced a set of ``scalar harmonics'', ${\cal Y}^\Sigma$, which are any set of complete functions on the internal manifold, labelled by the index $\Sigma$, which throughout will be raised/lowered by $\delta_{\Sigma\Omega}$. In our setup the internal space is topologically $S^5$ and we can choose this set of functions to be the scalar harmonics on the round $S^5$. Note that this can be done even though the metric of the 10d Pilch-Warner solution preserves only $\SU(2) \times \U(1) \subset \SO(6)$ as isometries. Finally, the KK fluctuations for the spin-1, tensor, and spin-2 fields are denoted by $A_\mu{}^{\uA,\Sigma}$, $B_{\mu\nu,\uA,\Sigma}$, and $h_{\mu\nu,\Sigma}$, respectively. Moreover, the KK fluctuations of the 42 scalar fields are represented by $j_{I,\Sigma}$, $I = 1, \ldots, 42$, which for fixed $\Sigma$ is an algebra coset element $j_{I,\Sigma} \in \mathfrak{e}_{6(6)} \ominus \mathfrak{usp}(8)$ appearing under the projection ${\cal P}_{\uA\uB,I}$ to ensure that $\gM_{MN} \in {\rm E_{6(6)}}/\USp(8)$.

Plugging the fluctuation Ansatz \eqref{eq:FluctAnsatz} into the linearized ExFT equations of motion immediately yields the mass matrices for the KK fields. The mass matrices are expressed in terms of the embedding tensor $X_{\uM\uN}{}^{\uP}$ of the 5d supergravity, the scalar fields ${\cal V}_{\uM}{}^{\uA}$ defining the PW vacuum in the 5d theory, and the linear action of the twist matrix on the harmonics, ${\cal T}_{\uM\,\Sigma}{}^{\Omega}$, defined as
\begin{equation} \label{eq:TMatrices}
	\gL_{\cU_{\uM}} {\cal Y}^\Sigma = - {\cal T}_{\uM}{}^{\Sigma}{}_{\Omega}\, {\cal Y}^\Omega \,.
\end{equation}
Whenever the gauge group of the gauged supergravity is compact, as it is in this case, the definition \eqref{eq:TMatrices} reduces to
\begin{equation}
	L_{{\cal K}_{\uM}} {\cal Y}^\Sigma = - {\cal T}_{\uM}{}^{\Sigma}{}_{\Omega}\, {\cal Y}^\Omega \,,
\end{equation}
in terms of the ordinary Lie derivative $L$ of Killing vectors ${\cal K}_{\uM}$ on the internal manifold, which can be read off from the twist matrix, $U_{\uM}{}^{M}$. One can show that the ${\cal T}_{\uM}$ matrices are generators of the gauge group of the gauged supergravity, in the representation of the scalar harmonics. When presenting the explicit formulae for the mass matrices of the fluctuations above, the embedding tensor and the ${\cal T}_{\uM}$ matrices appear dressed with the scalar matrices as follows
\begin{equation}
	\begin{split} \label{eq:DressedObj}
		X_{\uA\uB}{}^{\uC} &= \left(\mathcal V^{-1}\right)_{\uA}{}^{\uM}\,\left(\mathcal V^{-1}\right)_{\uB}{}^{\uN}\, X_{\uM\uN}{}^{\uP}\, \cV_{\uP}{}^{\uC} \,, \\
		{\cal T}_{\uA}{}^{\Sigma}{}_{\Omega} &= \left(\mathcal V^{-1}\right)_{\uA}{}^{\uM}\, {\cal T}_{\uM}{}^{\Sigma}{}_{\Omega} \,.
	\end{split}
\end{equation}

The spin-2 mass matrix is
\begin{equation} \label{eq:MassSpin2}
	\mathbb{M}_{\Sigma\Omega}^{(2)} = - {\cal T}_{\uA,\Sigma\Lambda} {\cal T}_{\uA,\Lambda\Omega} \equiv - \left( {\cal T}_{\uA} {\cal T}_{\uA} \right)_{\Sigma\Omega} \,,
\end{equation}
where we have raised/lowered the $\uA$ and $\Sigma$ indices with $\delta_{\uA\uB}$ and $\delta_{\Sigma\Omega}$, respectively, a convention we will adopt henceforth. Thus, the spin-2 mass matrix takes the form of a ``generalized'' Casimir operator constructed from the ${\cal T}_{\uA}$ matrices.

Similarly, the tensor mass matrix is
\begin{equation}
	\mathbb{M}_{(t)}^{\underline{A}\Sigma,\underline{B}\Omega} =
	\frac{1}{\sqrt{10}}\left(
	-2\, d^{\uC\uD\uA}\, X_{\uC\uD}{}^{\uB}\,\delta_{\Sigma\Omega} + 10\,   d^{\underline{A}\underline{B} \underline{C}}\, {\cal T}_{\underline{C},\Sigma\Omega}
	\right)
	\,,
	\label{eq:MassTensor}
\end{equation}
and involves the symmetric cubic invariant of ${\rm E_{6(6)}}$, $d^{\uA\uB\uC}$, while the spin-1 mass matrix is given by
\begin{equation}
	\begin{split}
		\mathbb{M}^{(v)}_{\uA\Sigma,\uB\Omega}&=
		\frac16\,
		\,
		X_{\underline{AD}}{}^{\uC}\,
		\left(
		X_{\underline B\underline C}{}^{\underline D}
		+   
		X_{\underline B\underline D}{}^{\underline C}
		\right)
		\delta_{\Sigma\Omega}
		+
		\left(
		X_{\underline B\underline A}{}^{\underline C}
		+   
		X_{\underline B\underline C}{}^{\underline A}
		-X_{\underline{AB}}{}^{\uC}
		-X_{\underline{AC}}{}^{\uB}
		\right)
		{\cal T}_{\uC,\Sigma\Omega} \\
		&{} \quad 
		-6
		\left(
		\mathbb{P}_{\uA}{}^{\uC}{}_{\uB}{}^{\uD}
		+\mathbb{P}_{\uC}{}^{\uA}{}_{\uB}{}^{\uD}
		\right)
		\left( {\cal T}_{\uC}  {\cal T}_{\uD}\right)_{\Sigma\Omega}
		\,,
		\label{eq:MassVec}
	\end{split}
\end{equation}
where $\mathbb{P}_{\uA}{}^{\uB}{}_{\uC}{}^{\uD}$ is the projector onto the adjoint of ${\rm E}_{6(6)}$. Finally, the scalar mass matrix is
\begin{equation}
	\begin{split}
		\mathbb{M}^{(s)}_{I\Sigma,J\Omega}
		&=
		X_{\uA\uE}{}^{\uF} X_{\uB\uF}{}^{\uE} \;
		{\cal P}_{\uA\uD}{}^I\, {\cal P}_{\uB\uD}{}^J\, \delta^{\Sigma\Omega}
		\\
		&\quad + \frac15
		\left( X_{\uA\uE}{}^{\uF} X_{\uB\uE}{}^{\uF} + X_{\uE\uA}{}^{\uF} X_{\uE\uB}{}^{\uF} + X_{\uE\uF}{}^{\uA} X_{\uE\uF}{}^{\uB}  \right)
		{\cal P}_{\uA\uD}{}^I\, {\cal P}_{\uB\uD}{}^J\, \delta^{\Sigma\Omega}
		\\
		&\quad+  \frac25
		\left( X_{\uA\uC}{}^{\uE} X_{\uB\uD}{}^{\uE} - X_{\uA\uE}{}^{\uC} X_{\uB\uE}{}^{\uD} - X_{\uE\uA}{}^{\uC} X_{\uE\uB}{}^{\uD} \right) 
		{\cal P}_{\uA\uB}{}^I\, {\cal P}_{\uC\uD}{}^J\, \delta^{\Sigma\Omega}
		\\
		& \quad
		-4\,
		X_{\underline{AC}}{}^{\underline{D}}\, {\cal T}_{\underline{B},\Omega\Sigma}
		\, {\cal P}_{\uA\uB}{}^I {\cal P}_{\uC\uD}{}^J
		-4\,X_{\underline{CA}}{}^{\underline{B}} \, {\cal T}_{\underline{C},\Omega\Sigma}
		\, {\cal P}_{\uA\uD}{}^I\, {\cal P}_{\uB\uD}{}^J \\
		& \quad
		+ 12\, {\cal T}_{\underline{A},\Omega\Lambda}
		{\cal T}_{\underline{B},\Lambda\Sigma}\,
		{\cal P}_{\uA\uD}{}^I\, {\cal P}_{\uB\uD}{}^J
		-  {\cal T}_{\underline{C},\Omega\Lambda} {\cal T}_{\underline{C},\Lambda\Sigma} \,
		{\cal P}_{\uA\uB}{}^I\, {\cal P}_{\uA\uB}{}^J
		\,.
		\label{eq:MassScalar}
	\end{split}
\end{equation}

\subsubsection{Further simplifying the mass matrices}

Written as in \eqref{eq:FluctAnsatz}, the KK fluctuations live in the tensor product of the maximal 5d supergravity with the scalar fluctuations on the internal space. However, not all these fluctuations are physical. Some of the massless scalar fields amongst the $j_{I,\Sigma}$ are Goldstone modes for massive vector fields. Similarly, some massive vector and scalar fields are eaten by massive spin-2 fields, massless vector fields are eaten by the massive tensor fields, whilst the massless tensor fields are unphysical. This can be used to further simplify the mass matrices in \eqref{eq:MassTensor}, \eqref{eq:MassVec}, and \eqref{eq:MassScalar}.

Following \cite{Guarino:2020flh}, we can define the projection onto Goldstone scalars
\begin{equation}
	\Pi_{\uA\Sigma,I\Omega} = \delta_{\Sigma\Omega}\, X_{\uA\uC}{}^{\uD}\, {\cal P}_{\uC\uD,I} - 6\, {\cal P}_{\uA\uD,I}\, {\cal T}_{\uD,\Omega}{}^\Sigma \,.
\end{equation}
Using this projection matrix, the scalar mass matrix can be written in the more compact form
\begin{equation}
	\begin{split} \label{eq:MassScalarSimple}
		\mathbb{M}^{(s)}_{I\Sigma,J\Omega} &= {\cM}_{IJ}^{(0)}\, \delta_{\Sigma\Omega} + \delta_{IJ}\, \mathbb{M}^{(2)}_{\Sigma\Omega} + {\cal N}_{IJ}{}^{\uC}\, {\cal T}_{\uC,\Sigma\Omega} - \frac13 \left( \Pi^T \Pi \right)_{I\Sigma,J\Omega} \,,
	\end{split}
\end{equation}
in terms of the spin-2 mass matrix, $\mathbb{M}^{(2)}_{\Sigma\Omega}$, given in \eqref{eq:MassSpin2}, the 5-dimensional supergravity mass matrix for the scalar fields $\cM_{IJ}^{(0)}$ and the matrix
\begin{equation}
	{\cal N}_{IJ}{}^{\uC} = - 4 \left( X_{\uC\uA}{}^{\uB} + 6 X_{\uA\uB}{}^C \right) {\cal P}_{\uA\uD}{}^{[I}\, {\cal P}_{\uB\uD}{}^{J]} \,.
\end{equation}
Note that the final term in \eqref{eq:MassScalarSimple} induces a shift that only affects the Goldstone modes.

We can apply the same logic to simplify the vector and tensor mass matrices. We define the projection onto the vector fields that are eaten by massive spin-2 fields as
\begin{equation}
	\Pi_{\uA\Sigma,\uB\Omega} = \left( {\cal T}_{\uA} {\cal T}_{\uB} \right)_{\Sigma\Omega} \,.
\end{equation}
Since massless vector/tensors are unphysical and absorbed by the massive tensors/vectors, we consider the sum of the vector and the square of the tensor mass matrix. After some manipulations, this gives rise to the compact expression
\begin{equation} \label{eq:MassVectorSimple}
	\mathbb{M}_{\uA\Sigma,\uB\Omega}^{(v)} - \left( \mathbb{M}_{(t)}^2 \right)^{\uA\Sigma,\uB\Omega} = M_{\uA\uB}^{(0)} + {\cal N}_{\uC,\uA\uB} {\cal T}_{\uC,\Sigma\Omega} + \delta_{\uA\uB} \mathbb{M}^{(2)}_{\Sigma\Omega} - \frac53 \Pi_{\uA\Sigma,\uB\Omega} \,,
\end{equation}
where $M_{\uA\uB}^{(0)}$ is the combined vector/tensor mass matrix in 5d gauged supergravity and we defined
\begin{equation}
	{\cal N}_{\uC,\uA\uB} = X_{\uC\uA}{}^{\uB} - X_{\uC\uB}{}^{\uA} - 6\, \mathbb{P}^{\uA}{}_{\uB}{}^{\uD}{}_{\uE}\, X_{\uE\uD}{}^{\uC} \,.
\end{equation}
Once again, the final term in \eqref{eq:MassVectorSimple} only affects the masses of vectors that are eaten by massive spin-2 fields, while the first three terms give the mass matrix whose eigenvalues determine the masses of the physical vector and tensor fields.

\subsubsection{KK spectroscopy for the PW solution}

Let us now give the relevant expressions for the embedding tensor, scalar harmonics and ${\cal T}_{\uM,\Sigma}{}^{\Omega}$ matrices needed to compute the mass matrices for all KK fluctuations around the 10d PW solution. We use the decomposition ${\rm E}_{6(6)} \longrightarrow \SL(6) \times \SL(2)$ in which the fundamental $\mbf{27}$ of ${\rm E_{6(6)}}$ decomposes into
\begin{equation}
	\mbf{27} \longrightarrow \left(\mbf{15},\mbf{1}\right) \oplus \left(\mbf{6}',\mbf{2}\right) \,.
\end{equation}
The only non-zero components of the $\SO(6)$ embedding tensor are then given by
\begin{equation}
	X_{\underline{MN}}{}^{\underline{P}} =
	\left\{
	\begin{array}{l}
		X_{ij,kl}{}^{mn} 
		= \mathtt g\,\sqrt{2}\,\delta_{[i}^{[m}\delta^{\phantom{m}}_{j][k}\delta_{l]}^{n]}\,, 	\\
		X_{ij}{}^{k\alpha}{}_{l\beta} =
		-\frac{\mathtt g}{\sqrt{2}} \delta_{[i}^{k}  \delta^{\phantom{k}}_{j]l}\,\delta_\beta^\alpha  \,.
	\end{array}
	\right.
	\label{eq:S5X}
\end{equation}

As discussed in more detail in \cite{Malek:2020yue}, we can use the scalar harmonics of the round $S^5$ as our complete set of functions because the consistent truncation is defined on $S^5$. For this, we use the embedding coordinates ${\cal Y}^i: S^5 \hookrightarrow \mathbb{R}^6$ satisfying
\begin{equation}
	{\cal Y}^i\, {\cal Y}^j\, \delta_{ij} = 1 \,.
\end{equation}
The $S^5$ scalar harmonics are now given by traceless polynomials of the ${\cal Y}^i$, i.e.
\begin{equation} \label{tower}
	{\cal Y}^\Sigma = \left\{ 1,\, {\cal Y}^i,\, {\cal Y}^{i_1i_2},\, \ldots,\, {\cal Y}^{i_1\ldots i_n}, \ldots \right\} \,,
\end{equation}
where we denote by ${\cal Y}^{i_1 \ldots i_n} \equiv {\cal Y}^{(\!(i_1} \ldots {\cal Y}^{i_n)\!)}$ the $n$-fold traceless symmetric polynomial. Thus, the $\Sigma$ index labels the $n$-fold symmetric traceless representation of $\SO(6)$ with Dynkin labels $\left[n,0,0\right]$.

Finally, the ${\cal T}_{\uM,\Sigma}{}^{\Omega}$ matrices are given by \cite{Malek:2020yue}
\begin{equation}
	\begin{split}
		{\cal T}_{\uM,i_1 \ldots i_n}{}^{j_1\ldots j_n} = n\, {\cal T}_{\uM,(\!(i_1}{}^{(\!(j_1} \delta_{i_2}^{j_2} \ldots \delta_{i_n)\!)}^{j_n)\!)} \,,
	\end{split}\label{eq:S5Tn}
\end{equation}
with
\begin{equation}
	{\cal T}_{\underline{M},k}{}^{l} = \left\{ \begin{array}{l}
		{\cal T}_{ij,k}{}^{l} = \frac{\mathtt g}{\sqrt{2}}\,\delta^{\phantom{ij}}_{k[i}\delta_{j]}^{l} \,, \\
		{\cal T}^{i\alpha}{}_{k}{}^{l} = 0 \,.
	\end{array}
	\right.
	\label{eq:S5T}
\end{equation}
Here we use the following summation convention for the harmonic indices $\Sigma$, $\Omega$
\begin{equation}
	A^\Sigma\, B_\Sigma = A\, B + A^i\, B_i + A^{i_1i_2}\, B_{i_1i_2} + \ldots + A^{i_1 \ldots i_n}\, B_{i_1\ldots i_n} + \ldots \,.
\end{equation}

Armed with the expressions \eqref{eq:S5X} and \eqref{eq:S5Tn}, as well as the scalar matrix \eqref{eq:5dGenMetric} for the PW solution \eqref{eq:PilchWarner}, we can now evaluate the mass matrices for the KK fluctuations \eqref{eq:MassSpin2}, \eqref{eq:MassTensor}, \eqref{eq:MassScalarSimple}, and \eqref{eq:MassVectorSimple}.

\subsection{The full KK spectrum}
\label{subsec:KKspecExFT}

We now present the results from an explicit evaluation of the mass formulae in the previous section. We organize the calculation by KK level, $n$, i.e. by the degree of the polynomial in \eqref{tower}. For each value of $n$ we then calculate the mass matrices of all bosonic fields in the supergravity theory and determine their eigenvalues. To organize the results we map these masses into conformal dimensions of the LS SCFT using the holographic dictionary summarized below
\begin{equation}
\begin{split}
\text{spin-}2: \quad \Delta &= 2 + \sqrt{4+m^2 L_{}^2} \,,\\
\text{vectors}: \quad \Delta &= 2 + \sqrt{1+m^2 L_{}^2}\,,\\
\text{two-forms}: \quad \Delta &= 2 +\left| m L_{} \right|\,,\\
\text{scalars}: \quad \Delta &= 2 \pm \sqrt{4+m^2 L_{}^2}\,,
\end{split}
\end{equation}
where $L$ is the IR AdS length scale in \eqref{eq:LUVIR}. We then arrange these results into 4d $\mathcal N=1$ superconformal multiplets that we denote as
\begin{equation}
X\bar Y[\Delta;j_1,j_2;r]\otimes [k]\,,
\end{equation}
where $[\Delta;j_1,j_2;r]$ denote the conformal dimension, the Lorentz spin, and the $\U(1)_R$ charge, respectively, of the superconformal primary operator in a given multiplet. The label $[k]$ denotes the $\SU(2)_F$ spin. Finally, the letters $X$ and $Y$ indicate the specific 4d $\mathcal{N}=1$ superconformal multiplets. We use the notation of \cite{Cordova:2016emh} and summarize some of the pertinent details of the superconformal multiplet structure in Appendix~\ref{App: N=1 d=4 multiplets}. We note that in this work we have restricted to the calculation of the KK spectrum of the bosonic fields of type IIB supergravity. The spectrum of the fermionic excitations is uniquely determined by the structure of the superconformal multiplets summarized in Appendix \ref{App: N=1 d=4 multiplets}. 

At KK level $n=0$ we have the spectrum of excitations of the fields in the 5d $\mathcal{N}=8$ gauged supergravity theory. This was already computed in \cite{Freedman:1999gp}, and we find agreement with their results. For completeness we summarize this spectrum here. We find two short multiplets
\begin{equation}
\begin{aligned}
 {A_1\bar A_1}[3;\tfrac{1}{2},\tfrac{1}{2};0] \otimes  [0]\,, \quad \text{and} \quad A_2\bar{A}_2[2;0,0;0] \otimes  [1] \,,
\end{aligned}
\end{equation}
containing the energy-momentum tensor, and $\SU(2)_F$ flavor currents respectively. Additionally, at level $n=0$ we find ten semi-short multiplets and one long multiplet
\begin{equation}
\begin{aligned}
&{L\bar B_1}[\tfrac{9}{4};\tfrac{1}{2},0;\tfrac{3}{2}] \otimes  [\tfrac{1}{2}]  \,, \qquad \hspace{0.35cm} {A_1\bar L}[3;\tfrac{1}{2},0;0] \otimes  [0] \,, \qquad\hspace{0.3cm} L\bar{L}[1+\sqrt{7};0,0;0] \otimes  [0]\,, \\
& {B_1\bar L}[\tfrac{9}{4};0,\tfrac{1}{2};-\tfrac{3}{2}] \otimes  [\tfrac{1}{2}]    \,, \qquad {L\bar A_1}[3;0,\tfrac{1}{2};0] \otimes  [0]  \,,\\
&{L\bar B_1}[\tfrac{3}{2};0,0;1] \otimes  [1] \,,\qquad  \hspace{0.4cm}  {L\bar A_2}[\tfrac{11}{4};\tfrac{1}{2},0;\tfrac{1}{2}] \otimes  [\tfrac{1}{2}]  \,, \\
&{B_1\bar L}[\tfrac{3}{2};0,0;-1] \otimes  [1]   \,,\qquad   {A_2\bar L}[\tfrac{11}{4};0,\tfrac{1}{2};-\tfrac{1}{2}] \otimes  [\tfrac{1}{2}]  \,, \\
&{L\bar B_1}[3;0,0;2] \otimes  [0] \,,\\
&{B_1\bar L}[3;0,0;-2] \otimes  [0]  \,.
\end{aligned}
\end{equation}

We can now proceed systematically and extend these results to higher KK levels, $n>0$. The calculation becomes unwieldy even for small values of $n$. We summarize the explicit results up to level $n=3$ in Appendix~\ref{App: ExFT results}. Fortunately, there is a more compact presentation of the spectrum for general values of $n$. One can arrange all KK modes into superconformal multiplets and then for each family of such multiplets write down a generating function. It proves convenient to organize these generating functions in terms of the Lorentz spin of the superconformal primary operator in the superconformal multiplets. Each of these generating functions depends on the 4 ``fugacities'' summarized in Table~\ref{Tab: fugacities and labels}. We emphasize that the fugacities $(\nu,\rho,\kappa,\gamma)$ and the corresponding labels of the states $(n,r,k,p+2y)$ do not necessarily correspond to symmetries and charges in the LS theory but are introduced as a convenient mnemonic device. The combination $ p + 2 y$ associated to the fugacity $\gamma$ 
deserves a little more explanation. The label $p$ is the charge under $\U(1)_P\subset \SU(2)_2$, see the discussion around \eqref{eq:TRIR}, generated by
\begin{equation}
P \equiv \frac12 \left( T_1 + T_2 \right)\,.
\end{equation}
The label $y$ denotes  the $\U(1)_Y$ charge of the type IIB supergravity field from which a given KK mode arises. These charges can be found in \cite{Gunaydin:1984fk,Kim:1985ez,DHoker:2002nbb}. It is an interesting fact that although $\U(1)_P$ and $\U(1)_Y$ are not symmetries of the PW solution or the LS SCFT, and thus do not define actual quantum numbers, they can be used to organize the KK spectrum efficiently.

\begin{table}
\centering
\begin{tabular}{c|cccc}
\hline 
Fugacity & $\nu$ & $\rho$ & $\kappa$ & $\gamma$ \\ 
\hline 
Label of state & $n$ & $r$ & $k$ & $p+2y$ \\ 
\hline 
Interpretation & KK level & $\U(1)_R$ charge & $\SU(2)_F$ rep.& $\U(1)_P \times \U(1)_Y$ charge  \\ 
\hline 
\end{tabular} 
\caption{List of all fugacities used in the generating functions along with their associated labels and interpretation.}\label{Tab: fugacities and labels}
\end{table}

Superconformal multiplets with a spin-$1$ primary are of the form $X\bar Y[\Delta;\tfrac{1}{2},\tfrac{1}{2};r]\otimes [k]$ and are counted by the generating function
\begin{equation}\label{Eq: generating function spin-2}
Z_1 =Z_1^{A_1 \bar A_1} + Z_1^{L\bar A_1}+ \bar Z_1^{A_1 \bar L}+Z_1^{L\bar L}\,,
\end{equation}
where 
\begin{equation}\label{Eq: spin-2 generating function}
\begin{aligned}
Z_1^{A_1 \bar A_1} +Z_1^{L\bar A_1}+Z_1^{A_1 \bar L} =& \, \frac{1-\nu^2}{(1- \nu \, \rho)( 1-\frac{\nu}{\rho})}\,,\\
Z_1^{L\bar L} =&\, \frac{\rho  \left(1-\nu ^2\right) (1-\nu  \sqrt{\frac{\kappa }{\gamma  \rho }}) \left(1-\nu  \sqrt{\gamma  \kappa  \rho
   }\right)-\rho}{\nu  \left(1-\frac{\nu }{\rho }\right) (1-\nu  \rho ) (1-\nu  \sqrt{\frac{\kappa }{\gamma  \rho }})
   \left(1-\nu  \sqrt{\gamma  \kappa  \rho }\right)}\, \,.
\end{aligned}
\end{equation}
To read off the spectrum of superconformal multiplets one needs to expand these generating functions in a Taylor series in the fugacities $(\nu,\rho,\kappa,\gamma)$. The powers of the fugacities $(\nu,\rho,\kappa,\gamma)$ determine the KK level $n$, R-charge $r$, the highest weight state of the $\SU(2)_F$ spin state $[k]$, and the value of $p+2y$ respectively. To obtain the full $\SU(2)_F$ character for a given representation one has to apply the Weyl character formula to the generating function
\begin{equation}
\chi(\kappa) = \frac{\kappa\, Z(\kappa) - Z(1/\kappa)}{\kappa -1}\,.
\end{equation}

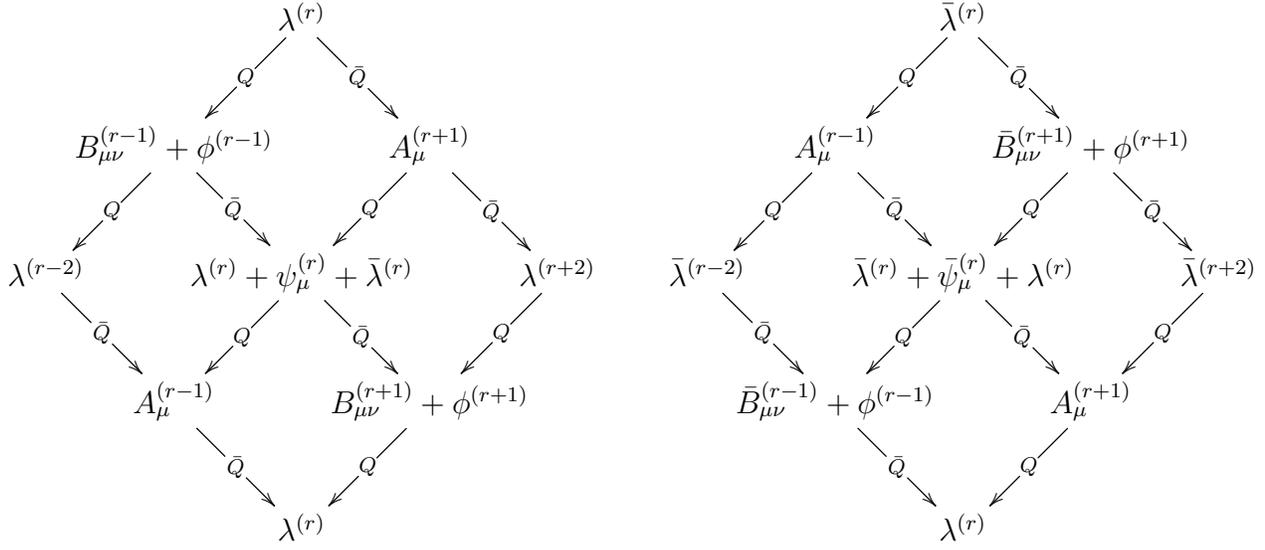
The superconformal multiplets with a spin-1/2 primary are of the schematic form $X\bar Y [ \Delta ; \tfrac{1}{2},0;r  ] \otimes [k]$, and $Y \bar X [ \Delta ; 0,\tfrac{1}{2};r  ] \otimes [k]$ and are counted by the generating functions
\begin{equation}\label{eq:genZ12}
Z_{1/2} = Z_{1/2}^{L\bar B_1} +  Z_{1/2}^{A_1 \bar L} +  Z_{1/2}^{L\bar A_2} + Z_{1/2}^{L\bar L} + \bar{Z}_{1/2}^{B_1\bar L} +  \bar{Z}_{1/2}^{L\bar A_1 } +  \bar{Z}_{1/2}^{A_2\bar L} + \bar{Z}_{1/2}^{L\bar L} \,,   
\end{equation}
where
\begin{equation}\label{eq:genZ12expl}
\begin{aligned}
Z_{1/2}^{L\bar B_1}  =&\, \frac{\gamma \rho \sqrt{\gamma\, \rho\, \kappa}}{1- \nu \sqrt{\gamma\, \rho\, \kappa}}\,,\qquad 
Z_{1/2}^{A_1 \bar L}  =\, \frac{\gamma +\frac{\nu}{\gamma \,\rho}}{1 - \frac{\nu}{\rho}}\,,\qquad 
Z_{1/2}^{L\bar A_2}  =\, \frac{\sqrt{\gamma\, \rho\, \kappa}}{1- \nu\sqrt{\gamma\, \rho\, \kappa}}\,,\\
Z_{1/2}^{L\bar L}  =&\, \nu  \left(\frac{\nu }{\gamma  \rho }+\gamma \right)\frac{ \kappa 
   (\frac{\gamma  (\rho -\nu )}{\gamma ^2 \rho +\nu }-\nu  (1-\nu  \rho
   ))+\sqrt{\frac{\kappa }{\gamma  \rho }}(1+\gamma  \rho ) (1-\nu  \rho ) +\rho
   }{(1-\frac{\nu }{\rho }) (1-\nu  \rho ) (1-\nu 
   \sqrt{\frac{\kappa }{\gamma  \rho }}) (1-\nu  \sqrt{\gamma  \kappa  \rho
   })} \,.
\end{aligned}
\end{equation}
The generating functions $\bar{Z}_{1/2}$ in \eqref{eq:genZ12} are obtained from the expressions in \eqref{eq:genZ12expl} by the substitution
\begin{equation}\label{Eq: complex conjugation generating functions}
\rho \rightarrow \rho^{-1}\,, \quad \text{and} \quad \nu \rightarrow \nu^{-1}\,.
\end{equation}

The superconformal multiplets with a spin-$0$ primary are of the form $X\bar Y[\Delta;0,0;r]\otimes [k]$ and are counted by the generating function
\begin{equation}\label{eq:genZ0}
Z_0 = Z_0^{A_2 \bar A_2} + Z_0^{L\bar B_1} + Z_0^{L\bar A_2} + \bar{Z}_0^{B_1\bar L} + \bar{Z}_0^{A_2\bar L} + Z_0^{L\bar L}\,,
\end{equation}
where
\begin{equation}\label{eq:genZ0expl}
\begin{aligned}
Z_0^{A_2 \bar A_2} =&\, \kappa\,,\qquad Z_0^{L\bar B_1} = \frac{\gamma\,\rho (\gamma \, \rho+\kappa)}{1-\nu  \sqrt{\gamma\, \rho\, \kappa}} \,,\qquad Z_0^{L\bar A_2} = \frac{\nu\,\sqrt{\gamma\, \rho\, \kappa} (\gamma \, \rho+\kappa)}{1-\nu  \sqrt{\gamma\, \rho\, \kappa}}\,,\\
Z_0^{L\bar L} =&\, \frac{1 + \nu(\nu + \rho \gamma^2 + \frac{1}{\rho \gamma^2}) + \kappa\, \nu(1- \nu^2)(\gamma+\frac{1}{\gamma}) + \kappa^2\, \nu^2 (1- \frac{\nu}{\rho})(1- \nu\,\rho)}{(1-\nu\, \rho)(1-\frac{\nu}{\rho})(1- \nu \sqrt{\kappa\, \rho\, \kappa})(1- \nu\sqrt{\frac{\kappa}{\gamma\, \rho}} )}\,.
\end{aligned}
\end{equation}
The generating functions $\bar Z_0$ in \eqref{eq:genZ0} are obtained from the expressions in \eqref{eq:genZ12expl} by again using the substitution in \eqref{Eq: complex conjugation generating functions}.

To completely specify the spectrum of superconformal multiplets we need to supplement the generating functions above with a formula for the conformal dimension of the superconformal primary operator in each multiplet. There is indeed a simple compact expression for these conformal dimensions in terms of the labels $(j_1,j_2,n,r,k,p+2y)$ that reads
\begin{align}
\Delta = 
1 + \sqrt{
7- 2 j_1 (j_1+1) - 2 j_2 (j_2+1) + \tfrac34 \left(r^2  - 2(p+2y)^2+ 2 \,{n} (n + 4) - 4 k (k+ 1)\right)}\,.
\label{DeltaAll}
\end{align}
In summary, the generating functions in \eqref{Eq: generating function spin-2}, \eqref{eq:genZ12}, and \eqref{eq:genZ0}, together with the conformal dimensions in \eqref{DeltaAll} completely determine the full spectrum of KK modes around the AdS$_5$ PW solution dual to the LS SCFT.

We were not able to rigorously derive the analytic expressions above in full generality. To obtain them we have instead used several complementary methods. The multiplets counted by the generating function in \eqref{Eq: generating function spin-2} contain a descendant operator with spin-2. The mass of this mode is determined by the spin-$2$ mass matrix in \eqref{eq:MassSpin2}. One can find a closed form expression for the eigenvalues of this mass matrix for any KK level $n$. The result is consistent with the expression in \eqref{DeltaAll} and reads:
\begin{equation}\label{Eq: spin2 from ExFT}
\Delta_{\text{spin-2}} = 
2 + \sqrt{
4 + \tfrac34 \left(r^2  - 2p^2+ 2 \,{n} (n + 4) - 4 k (k+ 1)\right)}\,.
\end{equation}
There is no dependence on the label $y$ in this expression since the spin-2 modes in type IIB supergravity come from the metric and are not charged with respect to $\U(1)_Y$.

As discussed in detail in Section~\ref{sec:IIB} below, the spin-2 spectrum can also be calculated directly  in type IIB supergravity by solving the scalar Laplace equation in the PW background. The result of this alternative calculation is the same as \eqref{Eq: spin2 from ExFT} and provides a non-trivial consistency check of the ExFT KK spectroscopy method. For spin-0 and spin-1 supergravity modes the KK spectrum results presented above has been checked  explicitly up to and including level $n=4$, while for the two-form excitations we have checked up to $n=5$. Yet another consistency check of the KK spectrum above can be performed by restricting to semi-short multiplets at arbitrary level $n$. These are precisely the multiplets that contribute to the superconformal index of the LS SCFT and, as we discuss in detail in Section~\ref{sec:index}, an explicit calculation of the index in the planar limit yields results that are in full agreement with the KK spectrum presented here.

To facilitate the comparison between the KK spectroscopy results and the superconformal index computation in Section~\ref{sec:index} we explicitly present here all KK towers of semi-short multiplets resulting from the analysis above.

From the spin-$1$ spectrum in \eqref{Eq: spin-2 generating function} we identify two towers of semi-short multiplets 
\begin{equation}
L\bar A_1 [\tfrac{6+3n}{2};\tfrac{1}{2},\tfrac{1}{2};n]\otimes [0]^{(0)}\,, \quad \text{and} \quad  A_1 \bar L [\tfrac{6+3n}{2};\tfrac{1}{2},\tfrac{1}{2};-n]\otimes [0]^{(0)}\,,
\end{equation}
where, for completeness, we have listed the value of $p+2y$ as a superscript on the $\SU(2)_F$ spin representation.\footnote{It is important to remember that the label $p+2y$ is not related to the symmetry of the LS theory and therefore there is no corresponding fugacity for it in the superconformal index.} At level $n=0$ these two towers degenerate to a single $A_1 \bar A_1[3;\tfrac{1}{2},\tfrac{1}{2};0]\otimes [0]^{(0)}$ multiplet  that contains the stress-energy tensor.

The spectrum in \eqref{eq:genZ12} contains eight towers of semi-short multiplets given by
\begin{equation}\label{Eq: spin12 semi-shorts}
\begin{aligned}
 L\bar{B}_1&[\tfrac{9+3n}{4};\tfrac{1}{2},0;\tfrac{n+3}{2}] \otimes  [\tfrac{n+1}{2}]^{([n+3]/2)}\,, \qquad \hspace{0.2cm}  B_1 \bar L[\tfrac{9+3n}{4},0,\tfrac{1}{2};-\tfrac{n+3}{2}] \otimes  [\tfrac{n+1}{2}]^{(-[n+3]/2)}\,,\\
 A_1 \bar L&[\tfrac{6+3n}{2};\tfrac{1}{2},0;-n] \otimes  [0]^{(1)} \,, \qquad\qquad \hspace{0.8cm}   L\bar A_1 [\tfrac{6+3n}{2};0,\tfrac{1}{2};n] \otimes  [0]^{(-1)} \,,\\
 A_1 \bar L&[\tfrac{6+3n}{2};\tfrac{1}{2},0;-n] \otimes  [0]^{(-1)} \,, \qquad\qquad \hspace{0.5cm}   L\bar A_1 [\tfrac{6+3n}{2};0,\tfrac{1}{2};n] \otimes  [0]^{(1)} \,,\\
L\bar{A}_2&[\tfrac{11+3n}{4};\tfrac{1}{2},0;\tfrac{n+1}{2}] \otimes  [\tfrac{n+1}{2}]^{([n+1]/2)} \,,\qquad \hspace{0.05cm} A_2 \bar L[\tfrac{11+3n}{4};0,\tfrac{1}{2};-\tfrac{n+1}{2}] \otimes  [\tfrac{n+1}{2}]^{(-[n+1]/2)} \,,
\end{aligned}
\end{equation}
where the multiplets in the second line have $n\geq 1$ and all other multiplets have $n\geq 0$.

The spectrum in \eqref{eq:genZ0} results in eight more towers of semi-short multiplets
\begin{equation}\label{eq:semishortspin0}
\begin{aligned}
 L\bar{B}_1&[\tfrac{6+3n}{4};0,0;\tfrac{n+2}{2}] \otimes  [\tfrac{n+2}{2}]^{([n+2]/2)}\,, \qquad \hspace{0.2cm}  B_1 \bar L [\tfrac{6+3n}{4};0,0;-\tfrac{n+2}{2}] \otimes  [\tfrac{n+2}{2}]^{(-[n+2]/2)}\,,\\
 L\bar{B}_1&[\tfrac{12+3n}{4};0,0;\tfrac{n+4}{2}] \otimes  [\tfrac{n}{2}]^{([n+4]/2)}\,, \qquad \hspace{0.45cm}  B_1 \bar L [\tfrac{12+3n}{4};0,0;-\tfrac{n+4}{2}] \otimes  [\tfrac{n}{2}]^{(-[n+4]/2)}\,,\\
 L \bar A_2&[\tfrac{8+3n}{4};0,0;\tfrac{n}{2}] \otimes  [\tfrac{n+2}{2}]^{(n/2)} \,, \qquad \qquad \hspace{0.3cm}  A_2 \bar L [\tfrac{8+3n}{4};0,0;-\tfrac{n}{2}] \otimes  [\tfrac{n+2}{2}]^{(-n/2)} \,,\\
 L \bar A_2&[\tfrac{14+3n}{4};0,0;\tfrac{n+2}{2}] \otimes  [\tfrac{n}{2}]^{([n+2]/2)} \,, \qquad \quad A_2 \bar L [\tfrac{14+3n}{4};0,0;-\tfrac{n+2}{2}] \otimes  [\tfrac{n}{2}]^{(-[n+2]/2)}  \,.
\end{aligned}
\end{equation}
At level $n=0$ the third line degenerates to a single $A_2 \bar A_2 [ 2;0,0;0 ]\otimes [1]^{(0)}$ multiplet which contains the $\SU(2)_F$ flavor current. The multiplets in the first three lines have $n\geq 0$, while the multiplets in the fourth line have $n\geq1$.

We note that in the spectrum above there are semi-short multiplets that contain marginal operators. Such marginal deformations, compatible with $\mathcal{N}=1$ supersymmetry, belong to $L\bar B_1 [3;0,0;2]$ and $B_1 \bar L [3;0,0;-2]$ multiplets. From the first line of \eqref{eq:semishortspin0} we find that at level $n=2$ we have the multiplets
\begin{equation}\label{eq:LBmarginal}
L\bar B_1 [3;0,0;2]\otimes [2]^{(2)}\,, \quad \text{and} \quad B_1 \bar L [3;0,0;-2]\otimes [2]^{(-2)}\,,
\end{equation}
while from the second line of \eqref{eq:semishortspin0} at $n=0$ we find
\begin{equation}
L\bar B_1 [3;0,0;2]\otimes [0]^{(2)}\,, \quad \text{and} \quad B_1 \bar L [3;0,0;-2]\otimes [0]^{(-2)}\,.
\end{equation}
We therefore find a total of 6 complex marginal deformations of the LS SCFT. To find the number of exactly marginal deformation, i.e. the dimension of the conformal manifold, we use the method of \cite{Green:2010da} and subtract the dimension of the $\SU(2)_F$ flavor symmetry. Therefore we conclude that the conformal manifold on which the LS SCFT resides has complex dimension $6-3=3$. It is expected that this conformal manifold is compact, see \cite{Perlmutter:2020buo} for a recent discussion.

We end our discussion on the KK spectrum with some comments on long multiplets. It is a notable feature of our results that there are infinite towers of KK modes dual to unprotected operators in the planar limit of the LS SCFT. This should be contrasted with the operator spectrum of $\mathcal{N}=4$ SYM where all KK modes are dual to protected operators. Moreover one may naively expect that the conformal dimensions of the unprotected multiplets are generic irrational numbers. From the expression for the conformal dimension in \eqref{DeltaAll} this indeed appears to be so. Note however that there are infinitely many long multiplets with rational conformal dimensions, one example of a class of such multiplets takes the form
\begin{equation}\label{eq:LLrational}
L\bar L [\tfrac{4+3n}{2} ; \tfrac{1}{2},\tfrac{1}{2};n-2] \otimes [1]^{(0)}\,, \quad n\geq2\,.
\end{equation}
A similar phenomenon was observed in the KK spectrum of the AdS$_5\times T^{1,1}$ type IIB solution and it was speculated in \cite{Ceresole:1999zs,Gubser:1998vd} that this may not be due to an accident but to some hidden symmetry of the Klebanov-Witten SCFT. Perhaps a similar phenomenon is at play here. Another possibility is that the multiplets in \eqref{eq:LLrational} are long multiplets that arise as a recombination of semi-short multiplets as one changes the exactly marginal couplings in the LS SCFT. To understand this better one has to analyze the operator spectrum along the full LS conformal manifold. Since the LS conformal manifold is strongly coupled and the dual AdS$_5$ supergravity solutions capturing the LS marginal deformations is not known, this analysis is unfortunately out of reach.

\section{The spin-2 spectrum from IIB supergravity}
\label{sec:IIB}

Calculating the full KK spectrum of the type IIB PW AdS$_5$ solution directly in the 10d supergravity is a daunting task which we do not know how to address. However, the spin-2 part of the spectrum is accessible by performing a calculation similar to the one done in~\cite{Klebanov:2009kp} for the AdS$_4$ CPW solution in 11d supergravity \cite{Corrado:2001nv}. This is due to the fact that the spectrum of gravitons can be reduced to that of a minimally coupled scalar field, which in turn can be explicitly calculated. We describe this calculation below and demonstrate explicitly that the result agrees with the spin-2 spectrum found using the ExFT techniques discussed in Section~\ref{sec:5dKK}.

\subsection{The 10d solution}
\label{subsec:PWsoln}

To set the stage we begin by presenting the full PW  AdS$_5$ solution of type IIB supergravity. The solution can be obtained by uplifting the 5d solution in Section~\ref{sec:5dKK} using the explicit uplift formulae presented in \cite{Baguet:2015sma}. We have performed this uplift explicitly and have confirmed that the resulting 10d background agrees with the one found in \cite{Pilch:2000ej}, see also \cite{Pilch:2000fu,Corrado:2001nv}. 

We start by choosing appropriate coordinates on $S^5$ that are adapted to the  $\SU(2)\times \U(1) \subset \SO(6)$ isometry of the PW solution. To this end we use the following three complex coordinates on $\mathbb{C}^3$  
\begin{equation}
z_1 = \mathcal Y^{1} + \rmi \mathcal Y^{4}\,, \quad z_2 = \mathcal Y^{2} + \rmi \mathcal Y^{3}\,,\quad w =  \mathcal Y^{5} - \rmi \mathcal Y^{6}\,.
\end{equation}
The $S^5$ is then defined by imposing the relation $\sum_{i=1}^{6}( \mathcal Y^{i})^2=1$ and can be parametrized by the following choice of angular coordinates.
\begin{equation}
z=\left(\begin{array}{c}
z_1 \\ 
z_2
\end{array} \right)
= \cos \theta \, R(\xi_1,\xi_2,\xi_3) \left(\begin{array}{c}
1 \\ 
0
\end{array} \right)\,, \qquad w = \rme^{-\rmi \phi} \sin \theta\,.
\end{equation}
Here we have defined the Euler rotation matrix
\begin{equation}
R(\xi_1,\xi_2,\xi_3) =  \rme^{\frac{\xi_2}{2}\rmi \sigma_3}\,  \rme^{\frac{\xi_1}{2}\rmi \sigma_1} \, \rme^{\frac{\xi_3}{2}\rmi \sigma_3}\,,
\end{equation}
with $\sigma_i$ the Pauli matrices. The ranges of the five angles are given by 
\begin{equation}
0\leq \xi_1\leq \pi\,, \quad  0\leq \xi_2\leq 2\pi\,, \quad 0\leq \xi_3 \leq 4\pi\,, \quad 0 \leq \theta \leq \frac{\pi}{2}\,, \quad 0\leq \phi \leq \pi\,.
\end{equation}
It proves useful to define also the coordinate
\begin{equation}
U = 1- |w|^2 = \cos^2 \theta\,. \label{eq:uDef}
\end{equation}
To write the metric on $S^5$ it is convenient to use a set of three one-forms, $\Theta_u$, $u = 1, 2, 3$, defined by
\begin{equation}
	R^{-1} \cdot \rmd R = \frac{\rmi}{2} \sigma_u \Theta_u\,,
\end{equation}
which have the explicit form
\begin{equation}\label{Eq: explicit one forms}
\begin{aligned}
\Theta_1 + \rmi \Theta_2 =\, \rme^{\rmi \xi_3}  (\rmd \xi_1 - \rmi \sin \xi_1 \rmd \xi_2)  \,, \quad \Theta_3 =\, \rmd \xi_3 + \cos \xi_1 \rmd \xi_2\,.
\end{aligned}
\end{equation}
The metric on the round $S^5$ takes the following explicit form in these coordinates
\begin{equation}
\rmd \Omega^2_{S^5} = \frac{4}{\mathtt g^2} \left[ \rmd \theta^2 + \sin^2 \theta\, \rmd \phi^2 + \tfrac{1}{4}\cos^2 \theta\left( \Theta_1^2 + \Theta_2^2 + \Theta_3^2 \right) \right]\,. 
\end{equation}
Note that we have included the prefactor $L^2_{\text{UV}}=\frac{4}{\mathtt g^2}$, see \eqref{eq:LUVIR}, which determines the scale of the AdS$_5\times S^5$ solution in terms of the 5d supergravity coupling constant $\mathtt g$. 

Using the coordinates above we can present the PW solution in a relatively compact form. The 10d metric is given by\footnote{We use the type IIB supergravity conventions summarized in \cite{Bobev:2018hbq} and have checked explicitly that the background below solves the 10d equations of motion.}
\begin{equation}\label{Eq: 10D metric LS}
\rmd s_{10}^2 = \frac{1}{\sqrt{g_s}} \left(\frac{1}{\sqrt{\Delta}}\rmd s_{1,4}^2 + \sqrt{\Delta}\, \rmd \Omega^2 \right)\,,
\end{equation}
where we have kept the string coupling $g_s$ explicit and the warp factor is given by
\begin{equation}
\Delta =\frac{3}{2^{5/3}(2-\cos^2\theta)}\,.
\end{equation}
By abuse of notation, we are using the same symbol $\Delta$ here as we are for the conformal dimensions elsewhere. Since this is the only occasion where the warp factor appears and since the context makes it clear which object is meant by $\Delta$, we hope this will not lead to confusion.
The metric on the deformed $S^5$ is
\begin{equation}\label{eq:S5sqPW}
\begin{aligned}
	\rmd \Omega^2 =& \frac{2^{4/3} ( 2-\cos^2 \theta)}{\mathtt g^2} \Bigg( \dd \theta^2 + \cos^2 \theta \left[ \frac{\Theta_1^2 + \Theta_2^2}{2(2-\cos^2 \theta)} + \frac{2 \Theta_3^2}{8-5 \cos^2 \theta} \right]\\
	&\qquad\qquad\qquad\qquad\qquad\qquad+ \sin^2 \theta \frac{(8-5\cos^2 \theta)}{3 (2-\cos^2\theta )^2} \left[ \frac{\cos^2 \theta}{8-5\cos^2 \theta} \Theta_3 + \dd \phi \right]^2\Bigg)\,.
\end{aligned}
\end{equation}
The NS-NS and R-R two-forms are combined into a complex two-form as
\begin{equation}
 B_2 + \rmi g_s C_2 =- \frac{\rme^{\rmi \phi} \cos\theta}{2 \mathtt g^2} \left( \Theta_1 + \rmi \Theta_2 \right)  \w \left( 2\, \rmd \theta + \rmi \frac{\sin 2\theta}{2- \cos^2 \theta} (\rmd \phi - \Theta_3)  \right)  \,.
\end{equation}
The R-R 4-form reads
\begin{equation}
C_4 =  \frac{2}{\mathtt g^4 g_s} \frac{\cos^4 \theta}{\cos^2 \theta -2} \Theta_1 \w \Theta_2 \w \Theta_3 \w \rmd \phi\,.
\end{equation}
Finally, the axion and dilaton are trivial
\begin{equation}
C_0 = 0\,, \qquad  \rme^{\Phi} = g_s\,.
\end{equation}
As described in great detail in \cite{Pilch:2000ej}, the full ten-dimensional solution is invariant under the $\SU(2)_F \times \U(1)_R$ isometries, and a combined $\U(1)_R \times \U(1)_Y $ action. To show this one has to take the $\U(1)_Y$ charges of the IIB fields into account, which we list in table \ref{Tab: IIB fields and U1Y charges}.  

Note that here we have presented the AdS$_5$ PW solution dual to the IR LS SCFT. The 10d gravitational description of the RG flow connecting $\mathcal{N}=4$ SYM with the LS SCFT is in terms of a domain wall solution of type IIB supergravity constructed in \cite{Pilch:2000fu}.

\begin{table}
\centering
\begin{tabular}{c|cccccc}
\hline 
IIB field & $g$ & $\Phi$ & $C_0$ & $B_2$ & $C_2$ & $C_4$  \\ 
\hline 
$\U(1)_Y$ charge & $0$ & $\pm2$ & $\pm2$ & $\pm1$ & $\pm1$ & $0$  \\ 
\hline 
\end{tabular} 
\caption{Type IIB supergravity fields corresponding to all bosonic KK modes, along with their $\U(1)_Y$ charges.}\label{Tab: IIB fields and U1Y charges}
\end{table}
%

\subsection{The spin-2 spectrum}
\label{subsec: spin2 spectrum}

The goal now is to study perturbation of the PW solution presented above which carry spin-2 on the non-compact AdS$_5$ part of the background and non-trivial internal quantum numbers under the $\SU(2)\times \U(1)$ isometry of the squashed sphere in \eqref{eq:S5sqPW}. To this end we follow the approach used in \cite{Klebanov:2009kp}.

The spin-2 fluctuations of interest can be written as a perturbation of the 5d metric of the form
\begin{equation}
g_{\mu\nu} = \hat{g}_{\mu\nu} + h_{\mu\nu}\,,
\end{equation}
where $\hat{g}$ denotes the background AdS$_5$ metric with coordinates $x^\mu = (x^0,x^1,x^2,x^3,r)$. In order to isolate the transverse-traceless part of the fluctuations, i.e. remove the spin-0 and spin-1 components of the tensor $h_{\mu\nu}$, we impose the conditions
\begin{equation}
h^{m}_{\phantom{m}m} = 0 \,, \quad \text{and} \quad \partial^m h_{mn} = 0\,, 
\end{equation}
where the Latin indices run over the 4d Minkowski space on the boundary of AdS$_5$ parametrized by $(x^0,x^1,x^2,x^3)$. Additionally we can choose a gauge such that 
\begin{equation}
h_{rm} = 0\,.
\end{equation}
As discussed in \cite{Gubser:1997yh,Constable:1999gb,Klebanov:2009kp}, the spectrum of these metric fluctuations is the same as the spectrum of a minimally coupled scalar field in the full 10d space-time. In other words, to find the spectrum of interest we have to solve the equations of motion for a scalar field, $\varphi$, with action
\begin{equation}
S = -\frac12\int \rmd ^{10} x \sqrt{-g_{10}}\, (\partial\varphi )^2 \,,
\end{equation}
where the 10d metric is given in \eqref{Eq: 10D metric LS}. To solve this scalar equation of motion we employ the separation of variables ansatz
\begin{equation}
\varphi = \Phi(x^m,r) Y(y^a)\,,
\end{equation}
where $y^a = (\xi_1,\xi_2,\xi_3,\theta,\phi)$ are the coordinates on the internal $S^5$. Using this the scalar equations of motion can be written as
\begin{equation}
 \square \varphi = Y(y^a) \square_5 \Phi(x^m,r) + \Phi(x^m,r) \mathcal L\left[ Y(y^a) \right] = 0\,,
\end{equation}
where $\square_5$ is the Laplacian on AdS$_5$ and we have defined the following differential operator on $S^5$
\begin{equation}\label{Eq: S5 differential}
\mathcal L = \frac{\Delta^{-1/2}}{\sqrt{-g_{10}}}\partial_a \left( \sqrt{-g_{10}}\, g^{ab}_{10} \,\partial_{b} \right)\,,
\end{equation}
not to be confused with the generalised Lie derivative briefly mentioned in section \ref{subsec:ExFT}. 
To find the mass spectrum of interest we need to find the eigenfunctions, $Y$, and eigenvalues of this differential operator 
\begin{equation}\label{Eq: eigenvalue problem scalar on sphere}
\mathcal L \, Y(y^a) = -m^2 Y(y^a) \quad \Rightarrow  \quad \square_5 \Phi(x^m,r) = m^2 \Phi (x^m,r)\,.
\end{equation}
Since the squashed $S^5$ has an $\SU(2)$ isometry it is useful to first find the eigenfunctions of the quadratic Casimir operator of $\SU(2)$. This differential operator is constructed from the 3 Killing vectors associated with this isometry as follows
\begin{equation}
\xi^u = z \cdot (\sigma^u)^T \cdot \partial_z-  \overline z \cdot ( \overline \sigma^u)^T \cdot \partial_{\overline z}\,,
\end{equation}
where $\sigma^u$ are the Pauli matrices. The quadratic Casimir can then be written as
\begin{equation}
\mathcal C_2 = \xi^u \xi^u\,.
\end{equation}
The eigenfunctions of this operator can be written as
\begin{equation}\label{eq:Jtsdef}
J_{t,s} (z,\overline z) = \zeta_{u_1 u_2 \ldots u_s v_1 v_2 \ldots v_t} \prod\limits_{k=1}^t z^{u_k} \prod\limits_{l=1}^s \overline z^{v_l}\,,
\end{equation}
where $\zeta_{u_1 u_2 \ldots u_s v_1 v_2 \ldots v_t}$  is a completely symmetric and traceless tensor
\begin{equation}
\zeta_{u_1 u_2 \ldots u_t v_1 v_2 \ldots v_s}=\zeta_{(u_1 u_2 \ldots u_t v_1 v_2 \ldots v_s)}\,,\quad \text{and}\quad \zeta_{w_1 \ldots w_l u_{l+1} \ldots u_t w_1 \ldots w_l v_{l+1} \ldots v_s} = 0\,.
\end{equation}
Notice that we have to remove all traces that arise by summing paired $z$ and $\overline{z}$ indices. With this at hand one then finds the associated eigenvalues
\begin{equation}\label{Eq: harmonic equation SU2}
\mathcal C_2\, J_{t,s}(z,\overline z) =  (t+s)(t+s+2)J_{t,s}(z,\overline z) \,.
\end{equation}
As expected the eigenvalue is determined by the combination $t+s$, however we have kept both the $t$ and $s$ labels since the remaining $\U(1)_R$ symmetry of the 10d background will distinguish between them. 

To find the full solution to the eigenvalue problem in \eqref{Eq: eigenvalue problem scalar on sphere} we have to dress the $J_{t,s}$ functions with a dependence on $w$ and $\bar w$. To do this we follow \cite{Klebanov:2009kp} and use the following ansatz 
\begin{equation}\label{Eq: ansatz for S5 solutions}
Y(y^a) = J_{t,s}(z,\overline z) w^{n_R} H(U)\,,
\end{equation}
with $U$ the combination defined in \eqref{eq:uDef}.
Plugging \eqref{Eq: ansatz for S5 solutions} into the first equation of \eqref{Eq: eigenvalue problem scalar on sphere} we find that the eigenvalue problem reduces to the following differential equation for $H(U)$
\begin{equation}\label{Eq: simple diff eq}
U(U-1) H'' + \left( (a_+ + a_- +1 )U - b \right) H' + a_+ a_- H = 0\,.
\end{equation}
This is a hypergeometric equation with coefficients 
\begin{equation}\label{Eq: coefficients of hypergeometric}
\begin{split}
b &= 2+s+t\,,\\
a_{\pm} &= \frac{b+n_R}{2} \pm \frac14 \sqrt{2^{2/3}\, 6 \frac{m^2}{\mathtt g^2} -2n_R(n_R+s-t)+ \frac{5\left(t+s\right)^2}{2} -2 t s + 4(b+2)}\,.
\end{split}
\end{equation}
Equation \eqref{Eq: simple diff eq} is therefore solved by hypergeometric functions as
\begin{equation}\label{eq:Husoln}
H(U)= {}_2F_1\left[a_-,a_+,b,U\right] \,.
\end{equation}
Using \eqref{eq:Husoln}, \eqref{Eq: ansatz for S5 solutions}, and \eqref{eq:Jtsdef} we arrive at the final explicit form of the eigenfunctions $Y(y^a)$ that solve the first equation in \eqref{Eq: eigenvalue problem scalar on sphere}.

We still need to ensure that the eigenfunctions just constructed are regular over the whole internal space. This amounts to imposing that the function
\begin{equation}\label{Eq: u function of harmonics}
w^{n_R} H(U) = \left( \sin\theta\,\rme^{-\rmi \phi} \right)^{n_R} \,_2F_1 (a_-,a_+,b,\cos^2 \theta)
\end{equation}
is regular at $\theta=0$. The behavior of \eqref{Eq: u function of harmonics} at $\theta \sim 0$ depends on the sign of $n_R$ and accordingly we have to study two separate cases.

Let us start by assuming that $n_R\geq 0$. At $\theta = 0$ the function in \eqref{Eq: u function of harmonics} then blows up as
\begin{equation}
\lim\limits_{\theta \rightarrow 0}  w^{n_R} H(U)  \propto \frac{n_R!}{\theta^{n_R}  \Gamma \left[ a_-\right] \Gamma \left[ a_+\right]}\,.
\end{equation}
To ensure that this limit is regular we impose that 
\begin{equation}
a_- = -q \,, \qquad q \in \mathbb{N}\,.
\end{equation}
Using this in \eqref{Eq: coefficients of hypergeometric} we find the following quantized mass spectrum
\begin{equation}
m^2 L_{\text{IR}}^2 = \frac{3}{16} \Big( 12 n_R^2  + (3s+t)(s+3t) +  32 q(q+t+s+2) + 24(t+s)  + 4 n_R(8+8 q +5 s +3t) \Big)\,.
\end{equation}
%


Now we focus on the case $n_R< 0$. At $\theta = 0$ the function in \eqref{Eq: u function of harmonics} has the following limit
\begin{equation}
\lim\limits_{\theta \rightarrow 0}  w^{n_R} H(U)  \propto \frac{|n_R|! \, \theta^{n_R}}{  \Gamma \left[- a_- + b\right] \Gamma \left[ -a_+ + b\right]}\,.
\end{equation}
To ensure regularity we need to set
\begin{equation}
a_- = -q + n_R\,,
\end{equation}
with $q \in \mathbb{N}$ and $n_R>-q$. This leads to the following expression for the mass spectrum
\begin{equation}
m^2 L_{\text{IR}}^2 = \frac{3}{16} \Big( 12 n_R^2  + (3s+t)(s+3t) +32 q(q+t+s+2) + 24(t+s)  - 4 n_R(8+8 q +5 s +3t) \Big)\,.
\end{equation}

In summary, we find that regularity of the eigenfunction $Y(y^a)$ is ensured whenever
\begin{equation}
a_- = -q\,,\quad \text{for} \quad n_R\geq 0\,, \quad \text{or} \quad a_- = n_R-q\,,\quad \text{for} \quad n_R < 0\,,
\end{equation}
where $q$ is a non-negative integer. The expression for the mass can be written in the following form valid for any value of $n_R$
\begin{equation}\label{Eq: mass formula LS}
m^2 L_{\text{IR}}^2 =  \frac{3}{16} \Big( 12 n_R^2  + (3s+t)(s+3t)  +32 q(q+t+s+2) + 24(t+s) + 4 |n_R|(8+8 q +5 s +3t) \Big)\,.
\end{equation}
This expression can be simplified by using the same labels as the ones in Section~\ref{subsec:KKspecExFT}
\begin{equation}
n = 2k + p - r + 2q\,, \quad k = \frac{t+s}{2} \,, \quad r = \frac{t-s}{2} - |n_R|\,, \quad p = \frac{t-s}{2}\,.
\end{equation}
The mass formula in \eqref{Eq: mass formula LS} then reduces to
\begin{equation}\label{Eq: spin2 masses}
\begin{aligned}
m^2 L_{\text{IR}}^2 =&\, \frac34\left[ 2 n(n+4) - 4 k(k+1) + r^2-2p^2  \right] \,.
\end{aligned}
\end{equation}
Upon translating this expression to a formula for the spin-2 conformal dimensions we find that the result is the same as the spin-2 spectrum in \eqref{Eq: spin2 from ExFT}.

\section{The LS superconformal index}
\label{sec:index}

The superconformal index, defined and studied in \cite{Romelsberger:2005eg,Kinney:2005ej}, see \cite{Rastelli:2016tbz} for a review, is an effective tool for studying 4d $\mathcal{N}=1$ SCFTs. It can be defined either as a supersymmetric partition function of the SCFT on the Euclidean manifold $S^1_\beta \times S^3$ or as a Witten index defined as a trace over the Hilbert space of the SCFT defined in radial quantization. Schematically, the $\cN=1$ index $\mc{I}$ is given by
\begin{align}
\mc{I} = \Tr (-1)^{\rm F} \e^{-\beta \delta} \e^{-\mu_i q_i}\,,
\end{align}
where $\rm F$ is the fermion number, $\delta = \{Q,Q^\dagger\}/2$, $\beta$ is the radius the $S^1$, and $\mu_i$ are chemical potentials for global symmetries which commute with the supercharge $Q$ and have conserved charges $q_i$. Similarly to the Witten index, $\mathcal{I}$ is independent of $\beta$ and receives contributions only from states with $\delta=0$.

For a general $\cN=1$ SCFT with Poincar\'e supercharges $Q_\alpha,\,\wt{Q}_{\dot \alpha}$, the superconformal algebra contains the following anti-commutators\footnote{The conjugate supercharges $Q^{\dagger}$ are defined to correspond to the superconformal charges $S$.}
\begin{align}\label{eq:QQdag}
\{Q_\alpha,\,Q^{\beta}{}^\dagger\} =\delta^\beta_\alpha \left( \Delta +\frac{3}{2}r\right)+ 2M_{\alpha}^\beta\,,\qquad \{\wt{Q}_{\dot \alpha},\,\wt{Q}^\dagger{}^{\dot \beta}\} = \delta^{\dot \beta}_{\dot \alpha} \left(\Delta-\frac{3}{2}r\right) + 2\wt{M}_{\dot\alpha}^{\dot\beta}\,,
\end{align}
where $\Delta$ is the conformal dimension, $M_\alpha^\beta$ and $\wt{M}_{\alpha}^{\dot\beta}$ are the generators of $\SU(2)\times \SU(2)$ rotations on S$^3$ with quantum numbers $j_1$ and $j_2$, and $r$ is the superconformal $\U(1)$ R-symmetry charge. 

One can now project onto states annihilated by the first or the second anti-commutators in \eqref{eq:QQdag} to define the ``left'' and the ``right'' superconformal indices 
\begin{equation}\label{eq:ILRdef}
\begin{split}
\mc{I}_{\rm L}&=\Tr (-1)^{\rm F} t^{2(\Delta+j_1)} y^{2j_2}\prod_I u_I^{q_I},\qquad \delta_{\rm L} = \Delta -2j_1+\frac{3}{2}r=0\,, \quad (t,y) = (\rme^{-\beta},\rme^{-\mathfrak{J}_1})\,,\\
\mc{I}_{\rm R}&=\Tr (-1)^{\rm F} t^{2(\Delta+j_2)}y^{2j_1}\prod_I u_I^{q_I},\qquad \delta_{\rm R} = \Delta- 2j_2 -\frac{3}{2}r=0\,, \quad (t,y) = (\rme^{-\beta},\rme^{-\mathfrak{J}_2})\,,
\end{split}
\end{equation}
where $\mathfrak J_i$ are the chemical potentials for the Lorentz symmetries, and $\prod_I u_I^{q_I}$ captures the contributions to the index from global symmetries $G_I$ with fugacities~$u_I$.  

We are interested in computing the index for the $\mathcal{N}=4$ SYM theory as well as the LS $\mathcal{N}=1$ SCFT. To this end it is useful to set up some notation about the symmetries in the problem and write the $\mathcal{N}=4$ theory in $\mathcal{N}=1$ language. The $\cN=4$ gauge multiplet breaks  into an $\cN=1$ vector multiplet, $V=(A,\,\lambda)$, consisting of the gauge field, $A$ and gaugino, $\lambda$, and three chiral multiplets $\Phi_i = (\vphi_i,\,\psi_i)$ built from Weyl fermions $\psi_i$ and complex scalars $\vphi_i$ in the adjoint representation of the gauge group. The superpotential of the $\mathcal N=4$ theory is the $m\rightarrow 0$ limit of
\begin{equation}\label{Eq: superpot sec2}
W = {\rm Tr}\, \Phi_1[\Phi_2,\Phi_3] + \frac{m}{2}\Phi_1^2\,.
\end{equation}
We have collected the charges and conformal dimensions of these elementary fields, or ``letters'', in  Table~\ref{tab:UV-letters}. We now proceed to elaborate on the notation used in this table. The UV superconformal $\U(1)$ R-symmetry is defined in \eqref{eq:TRUV} in terms of the three block-diagonal Cartan generators of the $\SO(6)$ R-symmetry of the $\mathcal{N}=4$ theory. It is also convenient to use another basis, $R_i$, for the Cartan subalgebra of $\SO(6)$ employed in the index calculations in \cite{Kinney:2005ej}. In this basis the UV $\U(1)$ superconformal R-symmetry generator reads
\begin{align}
r{}^{\rm UV} = \frac{1}{3}(3R_1+2R_2 +R_3)\,.
\end{align}
Turning on the superpotential mass term, i.e. taking $m\neq 0$ in \eqref{Eq: superpot sec2}, triggers the LS RG flow that ends in the $\mathcal{N}=1$ IR SCFT. The $\U(1)$ superconformal R-symmetry in the IR is given by the linear combination of $T_i$ as in \eqref{eq:TRUV} and can be written in the $R_i$ basis as
\begin{align}
r^{\rm IR} = \frac{1}{2}(2R_1 + R_2)\,.
\end{align}
The LS $\mathcal{N}=1$ SCFT enjoys an $SU(2)_F$ flavor symmetry which rotates the chiral superfields $\Phi_{2,3}$ and has a Cartan generator given by
\begin{align}
\tilde{R}_{F}= \frac{1}{2}(T_1 - T_2)\,, \qquad \tilde{R}_F = -\frac{1}{2}R_2\,.
\end{align}
%

\begin{table}[ht]
\begin{center}
\begin{tabular}{|c || c | c | c | c | c | c | c|} \hline
Field & $(-1)^{\rm F}(\Delta^{\rm{UV}},\,j_1,\,j_2)$ &$[T_1,\,T_2,\,T_3]$& $[R_1,\,R_2,\,R_3]$ & $r^{\rm UV}$ & $\Delta^{\rm{IR}}$ & $r^{\rm{IR}}$ & $\tilde{R}_F$ \\\hline
$\vphi_1$ & $(1,0,0)$ & $[0,0,1]$& $[1,0,-1]$ & $\frac{2}{3}$ & $\frac{3}{2}$  & --- & --- \\\hline
$\vphi_2$ & $(1,0,0)$ & $[0,1,0]$ & $[1,-1,1]$ &$\frac{2}{3}$& $\frac{3}{4} $ & $\frac{1}{2}$ &$-\frac{1}{2}$ \\\hline
$\vphi_3$ & $(1,0,0)$ & $[1,0,0]$& $[0,1,0]$ &$\frac{2}{3}$&  $\frac{3}{4}$& $\frac{1}{2}$ &$\frac{1}{2}$ \\\hline
$\psi_1$   &$-(\frac{3}{2},\pm \frac{1}{2},0)$ & $[-\frac{1}{2},-\frac{1}{2},\frac{1}{2}]$& $[0,0,-1]]$& $-\frac{1}{3}$& $2$& --- & ---\\\hline
$\psi_2$   & $-(\frac{3}{2},\pm \frac{1}{2},0)$& $[-\frac{1}{2},\frac{1}{2},-\frac{1}{2}]$ & $[0,-1,1]$ &$-\frac{1}{3}$&$\frac{5}{4}$&$-\frac{1}{2}$ & $-\frac{1}{2}$ \\\hline
$\psi_3$   &$-(\frac{3}{2},\pm \frac{1}{2},0)$ & $[\frac{1}{2},-\frac{1}{2},-\frac{1}{2}]$ & $[-1,1,0]$ &$-\frac{1}{3}$&$\frac{5}{4}$&$-\frac{1}{2}$ &$\frac{1}{2}$ \\\hline
$\lambda$  & $-(\frac{3}{2},0,\pm \frac{1}{2})$ &$[\frac{1}{2},\frac{1}{2},\frac{1}{2}]$ & $[1,0,0]$ &1&$\frac{3}{2}$& $1$&$0$\\\hline
$F$  & $(2,1,0)$ &$[0,0,0]$ &$[0,0,0]$ &0&$2$& $0$ &$0$\\\hline
$\partial_{\pm\pm}$  & $(1,\pm\frac{1}{2},\pm\frac{1}{2})$ &$[0,0,0]$ &$[0,0,0]$ &0&$1$& $0$ &$0$\\\hline
\end{tabular}
\caption{Charges and conformal dimensions for the ``letters'' in $\mathcal{N}=4$ SYM and the LS SCFT. $T_i$ and $R_i$ are the Cartan generators of $\SO(6)_R$ in two different basis, $r^{\rm UV}$ and $r^{\rm{IR}}$ are the generators under the UV and IR superconformal R-symmetry, respectively, while $\tilde{R}_F$ is the Cartan generator of the $\SU(2)_F$ flavor symmetry in the IR. Finally, $\Delta^{\rm{UV}}$ and $\Delta^{\rm{IR}}$ are the UV and IR conformal dimensions and $j_{1,2}$ are the Lorentz spin quantum numbers. }\label{tab:UV-letters}
\end{center}
\end{table}

After setting up the stage we are ready to proceed with the calculation of the superconformal index for the UV and IR theory. For concreteness we will present the results for the ``right'' index $\mathcal{I}_R$ and will omit the subscript from now on. For a QFT that admits a weak coupling limit on its conformal manifold, like $\mathcal{N}=4$ SYM, the calculation of the index amounts to enumerating the gauge invariant operators built out of the elementary letters which obey the shortening conditions in \eqref{eq:ILRdef}. For intrinsically strongly coupled SCFTs calculating the index is harder. However, in situations where the strongly coupled theory arises as an IR fixed point of a UV SCFT with a weakly coupled description one can take advantage of the invariance of the index under continuous deformations and compute it in the UV limit of the RG flow, see \cite{Romelsberger:2005eg,Romelsberger:2007ec} and \cite{Gadde:2010en} for a detailed explanation and justification for this procedure. Fortunately, the LS SCFT presents exactly such an example of an intrinsically strongly coupled theory and we employ the same approach as in \cite{Gadde:2010en} to compute its index. Since our main interest is to relate the information contained in the index to the supergravity spectrum calculations in the previous sections we will focus exclusively on the large $N$ limit of the gauge theory. In this limit one has to compute the so-called single letter index, see \cite{Gadde:2010en}, which we now proceed to discuss.

To begin, note that there is a universal contribution to every single letter index coming from acting with the derivative $\partial_{\pm \pm}$ an arbitrary number of times on any given letter.  From the information in Table~\ref{tab:UV-letters} we conclude that only $\partial_{\pm +}$ contributes to the right index.  Accounting for all of the charges we find that the contributions to the index from these derivatives are given by overall factors of the form
\begin{align}\label{eq:Idercontrib}
\sum_{i,j=0}^\infty(t^3y^{-1})^i(t^3y)^j= \frac{1}{(1-t^3y)(1-t^3/y)} \,.
\end{align}

We now examine the single letter index for a chiral multiplet of R-charge $r$ in a complex representation of the gauge and flavor symmetry group $\mc{R}$ with conjugate representation $\overline{\mc{R}}$. From the shortening condition $E = 2j_2+\frac{3}{2}r$, we see that only scalars $\vphi_i$ in $\mc{R}$ and chiral fermions $\overline{\psi}_i$ in $\overline{\mc{R}}$ contribute to the index.  Using this one finds the following compact expression for the chiral multiplet index
\begin{align}
\mathcal{I}_{\rm C}(U,V) = \frac{1}{(1-t^3 y)(1-t^3/y)}\left( t^{3r}\chi_{\mc{R}}(U,V) -t^{6-3r}\chi_{\overline{\mc{R}}}(U,V)\right)\,.
\end{align}
Here $U$ and $V$ are holonomy matrices for the gauge and flavor symmetries, respectively, and $\chi_{\mc{R}}$ is the character of the representation. Analogously, to study the vector multiplet index one can use that the chiral superfield $W^\alpha = \lambda^\alpha +\sigma^{\mu\nu}\theta^\alpha F_{\mu\nu} +\ldots$ has R-charge $+1$ along the RG flow. Furthermore,  the conformal dimension is fixed to $\frac{3}{2}$ both in the UV and IR. Therefore the single letter index for the vector multiplet is given by
\begin{align}\label{eq:IvecM}
\mathcal{I}_{\rm V}(U) = \frac{2 t^6 - t^3\left(y+\frac{1}{y}\right)}{\left(1-t^3 y\right)\left(1-t^3/y\right)}\chi_{\rm adj} (U) \,.
\end{align}
Here we have made use of the fact that all fields in the vector multiplet are in the adjoint representation of the gauge group.

With this at hand we are ready to present the full single letter index for the $\cN=4$ SYM theory as calculated in \cite{Kinney:2005ej}
\begin{align}\begin{split}\label{eq:N=4-SL-index}
\mathcal{I}^{\rm UV} &= \sum_{\rm letters} (-1)^{\rm F} t^{2(\Delta + j_2)}y^{2j_1} v^{R_2} w^{R_3}\\
&= \frac{t^2(v + w/v+1/w) - (t^3 y+t^3/y) -t^4(w +v/w + 1/v)+2t^6}{(1-t^3 y)(1-t^3/y)}\,,
\end{split}\end{align}
where $v,\,w$ are fugacities for the $R_{2,3}$ Cartan generators of $\SO(6)$. We have also used that the chiral multiplets are in the adjoint of the $\SU(N)$ gauge group and that there are no additional flavor symmetries in $\cN=4$ SYM.  

Equipped with the explicit expression for the single letter index we can follow the procedure in Section~5 of \cite{Gadde:2010en} to calculate the large $N$ limit of the index by a saddle point approximate. For the $\cN=4$ SYM index in \eqref{eq:N=4-SL-index} this was also done in \cite{Kinney:2005ej}. The result is
\begin{align}\label{eq:IIB-UV-index}
\mathcal{I}_{IIB}^{\rm UV} = \frac{t^2/w}{1-t^2/w} +\frac{t^2v}{1-t^2v} +\frac{t^2 w/v}{1-t^2w/v} - \frac{t^3/y}{1-t^3/y} - \frac{t^3y}{1-t^3y}.
\end{align}
This large $N$ result for the superconformal index is a prime target for a holographic comparison with the KK spectrum of type IIB supergravity on the AdS$_5\times S^5$ solution. Indeed the small $t$ expansion of $\mathcal{I}_{IIB}^{\rm UV}$ yields a power series of chiral primary operators of the schematic form $\Tr (\Phi^k)$ which can be mapped to the KK supergravity spectrum computed in \cite{Kim:1985ez,Gunaydin:1984fk}.

\subsection{The LS index}

We now move on to the calculation of the index for the LS theory of main interest here. We apply the R\"omelsberger prescription and compute the superconformal index of the LS theory by computing the UV index but using the charges in Table~\ref{tab:UV-letters} associated with the superconformal R-symmetry in the IR. We also refine the index by introducing a fugacity, $x$, for the $\SU(2)_F$ flavor symmetry. The superconformal index then takes the form
\begin{align}
\mc{I}_{\rm R} &= \Tr (-1)^{\rm F} t^{3(2j_2 +r^{\rm IR})} y^{2j_1} x^{2\tilde{R}_F}\,.
\end{align}

Note that using the data in Table~\ref{tab:UV-letters} we find that the scalar and fermion letters of the $\cN=1$ adjoint chiral $\Phi_1$ contribute to the single particle index as
\begin{equation}
\vphi_1:\quad t^3y^0x^0\,, \qquad\qquad\qquad \psi_1:\quad -t^3 y^0x^0\,,
\end{equation}
and therefore there is no contribution to the index from the $\Phi_1$ chiral multiplet. This is of course consistent with the fact that $\Phi_1$ is integrated along the RG flow from $\mathcal{N}=4$ SYM to the LS IR theory.

The 2 remaining chiral multiplets have the following contribution to the single letter index
\begin{equation}
\begin{split}
\Phi_2:&\quad t^{\frac{3}{2}} x -t^{\frac{9}{2}}x\,,\\
\Phi_3:&\quad t^{\frac{3}{2}} x^{-1} -t^{\frac{9}{2}}x^{-1}\,.
\end{split}
\end{equation}
As discussed around \eqref{eq:IvecM} the vector multiplet gives the same contribution to the index at the UV and IR fixed points. Combining these building blocks with the contribution from the derivatives given in \eqref{eq:Idercontrib} we arrive at the following expression for the single particle index of the LS SCFT
\begin{align}\label{eq:LS-single-particle-index}
\mathcal{I}^{\rm IR} = \frac{t^{\frac{3}{2}}(x+\frac{1}{x})-t^3(y+\frac{1}{y}) -t^{\frac{9}{2}}(x+\frac{1}{x})+2t^6}{(1-t^3y)(1-t^3/y)}\,.
\end{align}
Notice that this result for the LS index can be obtained from the index of the $\mathcal{N}=4$ SYM theory in \eqref{eq:N=4-SL-index} by the following substitution 
\begin{equation}\label{eq:wvrelUniv}
v\to 1/(x\sqrt{t})\,, \qquad\qquad w\to1/t\,.
\end{equation}
This relation between the UV and IR indices is not an accident and is due to the fact that the $\mathcal{N}=4$ SYM theory and the LS SCFT are related by the universal RG flow studied in \cite{Tachikawa:2009tt}. As discussed in Section~3 of \cite{Gadde:2010en} the superconformal indices of two theories related by such an RG flow are related.

We can now proceed to take the large $N$ limit of the single particle index \eqref{eq:LS-single-particle-index} as described in Section~5 of \cite{Gadde:2010en}. Alternatively we can obtain the same result by simply substituting the relations \eqref{eq:wvrelUniv} in \eqref{eq:IIB-UV-index}. The result is the following single-trace index of the LS theory
\begin{align}\label{eq:LS-IIB-single-trace-index}
\mathcal{I}^{\rm IR}_{IIB} = \frac{t^3}{1-t^3} +\frac{t^{\frac{3}{2}}/x}{1-t^{\frac{3}{2}}/x}+\frac{t^{\frac{3}{2}}x}{1-t^{\frac{3}{2}}x}-\frac{t^3y}{1-t^3y}-\frac{t^3/y}{1-t^3/y}\,.
\end{align}
This index should capture the information about the towers of KK modes dual to BPS operators in the AdS$_5\times\tilde{\text{S}}{}^5$ supergravity dual to the LS fixed point. Indeed, as we now proceed to show, this can be confirm by an explicit comparison with the supergravity KK spectrum calculations in the previous sections.

\subsection{KK spectroscopy from the index}

In order to compare the result for the index in \eqref{eq:LS-IIB-single-trace-index} with the KK spectroscopy calculation in supergravity we need to expand \eqref{eq:LS-IIB-single-trace-index} in an appropriate way which resembles the KK towers of short multiplets discussed in Section~\ref{subsec:KKspecExFT}. A similar calculation for the index of other $\mathcal{N}=1$ SCFTs with AdS$_5$ supergravity duals was discussed \cite{Nakayama:2006ur,Gadde:2010en}. One can show that \eqref{eq:LS-IIB-single-trace-index} can be expanded in the following way
\begin{equation}
\begin{split}\label{eq:LS-Index-IIB-3}
\mathcal{I}_{IIB}^{\rm IR} = &\Bigg[ \sum_{k=0}^\infty 2 t^{3(k+3)} -t^9- \sum_{k=0}^\infty t^{3(k+3)} \chi_{[1]}(y)+\sum_{k=0}^\infty t^{\frac{3(k+5)}{2}}\chi_{[1]}(y)\chi_{[k+1]}(x)\\&\quad-\sum_{k=1}^\infty t^{\frac{3(k+2)}{2}}\chi_{[1]}(y)\chi_{[k]}(x)+\sum_{k=2}^{\infty}t^{\frac{3k}{2}}\chi_{[k]}(x)-\sum_{k=1}^\infty t^{\frac{3(k+3)}{2}}\chi_{[k+1]}(x)\\&\quad\,\,+\sum_{k=0}^\infty t^{\frac{3(k+4)}{2}}\chi_{[k]}(x)-\sum_{k=0}^\infty t^{\frac{3(k+7)}{2}} \chi_{[k+1]}(x) \Bigg]\frac{1}{(1-t^3y)(1-t^3/y)}\\
&+ \frac{\chi_{[1]}(x)(t^{3/2}-t^{9/2}) -\chi_{[1]}(y)t^3 +2t^6}{(1-t^3y)(1-t^3/y)}.
\end{split}
\end{equation}
This expressions deserves some comments. First, we note that the last line in \eqref{eq:LS-Index-IIB-3} corresponds to a contribution to the index from a decoupled LS theory with a $\U(1)$ gauge group. The appearance of this free contribution is due to the fact that the results for the LS index in \eqref{eq:LS-single-particle-index} and \eqref{eq:LS-IIB-single-trace-index} are for the theory with an $\U(N)$ gauge group. To compare with the dual AdS$_5$ supergravity solution we have to use the planar limit of the $\SU(N)$ LS theory and thus we should remove the contribution from the decoupled $\U(1)$ sector. Second, we remind the reader that the denominator on the first line of \eqref{eq:LS-Index-IIB-3} accounts for the contribution of derivatives acting on single trace operators as discussed around \eqref{eq:Idercontrib}. Finally, we note that we have used the notation $\chi_{[k]}$ to denote the character of the $\SU(2)$ representation with Dynkin label $k$, i.e. $\chi_{[1]}(y)$ accounts for a contribution of a spin-1/2 fermion in space-time, while $\chi_{[k]}(x)$ denotes a contribution from the spin-$k/2$ representation of the $\SU(2)_F$ flavor symmetry.

We now proceed to interpret the terms in the first 3 lines of \eqref{eq:LS-Index-IIB-3} as contributions from towers of semi-short superconformal multiplets. We use the notation summarized in Appendix~\ref{App: N=1 d=4 multiplets} for the 4d $\mathcal{N}=1$ superconformal multiplets and add an additional label $\otimes[\tfrac{k}{2}]$ to denote the $\SU(2)_F$ flavor symmetry representation.

We begin with the towers of $L\bar{B}_1$ multiplets. The operators in these multiplets that contribute to the right index are the superconformal primary operators since they obey the shortening condition $\delta_{\rm R}=\Delta -2j_2 -\frac{3r}{2}=0$. For every such multiplet in the sums in \eqref{eq:LS-Index-IIB-3} there is a corresponding conjugate $B_1\bar{L}$ multiplet with, $j_1 \leftrightarrow j_2$, $r \leftrightarrow -r$, and the same  $\SU(2)_F$ quantum numbers which obeys the shortening condition $\delta_{\rm L}=\Delta -2j_1 +\frac{3r}{2}=0$ and contributes to the left index \eqref{eq:ILRdef}. The contributions of these multiplets are as follows:

\begin{enumerate}[label=(\roman*)]
\item The second sum on the second line of \eqref{eq:LS-Index-IIB-3}, $\sum_{k=2}^\infty t^{3k/2}\chi_{[k]}(x)$, corresponds to a tower of $L\bar{B}_1[\tfrac{3k}{4};0,0;\frac{k}{2}]\otimes [\tfrac{k}{2}]$ multiplets.

\item The first sum on the second line of \eqref{eq:LS-Index-IIB-3}, $-\sum_{k=1}^\infty t^{3(k+2)/2} \chi_{[1]}(y)\chi_{[k]}(x)$, corresponds to a tower of $L\bar{B}_1[\tfrac{3k+6}{4};\tfrac{1}{2},0;\frac{k+2}{2}]\otimes [\tfrac{k}{2}]$ multiplets.

\item The first sum on the third line of \eqref{eq:LS-Index-IIB-3}, $\sum_{k=0}^\infty t^{3(k+4)/2}\chi_{[k]}(x)$, arises from a tower of $L\bar{B}_1[\tfrac{3k+12}{4};0,0;\frac{k+4}{2}]\otimes [\tfrac{k}{2}]$ multiplets. 

\end{enumerate}

The remaining terms in \eqref{eq:LS-Index-IIB-3} are due to towers of $L\bar{A}_\ell$ ($\ell=1,2$) multiplets. Using the information in Appendix~\ref{App: N=1 d=4 multiplets} one can show that the superconformal primaries in such multiplets have $\delta_{\rm R}=2$ and thus do  not contribute to the right index. To understand which descendants contribute to the index we note that acting with a $Q$ supercharge shifts $\delta_{\rm R}$ as $\delta_{\rm R}\to \delta_{\rm R}+2$ while acting with a $\bar{Q}$ leads to $\delta_R \to \delta_R-2$. Therefore we conclude that the states in $L\bar{A}_\ell$ multiplets that contribute to the index are level 1  $\bar{Q}$ descendants. For every $L\bar{A}_\ell$ multiplet with $r\neq 0$ in the sums in \eqref{eq:LS-Index-IIB-3} there is a corresponding conjugate $A_\ell\bar{L}$ multiplet with, $j_1 \leftrightarrow j_2$, $r \leftrightarrow -r$, and the same  $\SU(2)_F$ quantum numbers which contains an operator with $\delta_{\rm L}=\Delta -2j_1 +\frac{3r}{2}=0$ and thus contributes to the left index \eqref{eq:ILRdef}. The contributions from the $L\bar{A}_\ell$ multiplets are as follows.

\begin{enumerate}
\item[(iv)]  The first two terms on the first line of \eqref{eq:LS-Index-IIB-3}, $ \sum_{k=0}^\infty 2t^{3(k+3)}-t^9$, capture the contribution of 2 towers of $L\bar{A}_1[\frac{3k+6}{2};0,\frac{1}{2};k]\otimes [0]$ multiplets with $k>0$. The $k=0$ term in the sum is subtle since there is actually only a single multiplet $L\bar{A}_1[3;0,\frac{1}{2};0]\otimes [0]$ multiplet and since it has $r=0$ there is no corresponding conjugate $A_1\bar{L}$ multiplet in the spectrum.
\item[(v)]  The third term on the first line of \eqref{eq:LS-Index-IIB-3}, $-\sum_{k=0}^\infty t^{3(k+3)}\chi_{[1]}(y)$, corresponds to a tower of $L\bar{A}_1[\frac{3k+6}{2};\frac{1}{2},\frac{1}{2};k]\otimes [0]$ multiplets.  Note that for $k=0$ the R-charge of the primary in the multiplet vanishes and thus there is no corresponding conjugate $A_1\bar{L}$ multiplet. In fact the multiplet with $k=0$ is the $A_1\bar{A}_1[3;\frac{1}{2},\frac{1}{2};0]\otimes [0]$ multiplet that contains the energy-momentum tensor.

\item[(vi)]  The final sum on the second line of \eqref{eq:LS-Index-IIB-3}, $-\sum_{k=1}^\infty t^{3(k+3)/2}\chi_{[k+1]}(x)$, corresponds to a tower of $L\bar{A}_2[\frac{3k+5}{4};0,0;\frac{k-1}{2}]\otimes [\frac{k+1}{2}] $ multiplets. Again, the multiplet with $k=1$ has a primary with R-charge 0 and thus there is no corresponding conjugate $A_2\bar{L}$ multiplet. More precisely, the multiplet with $k=1$ is an $A_2\bar{A}_2[2;0,0;0]\otimes [1] $ multiplet which contains the $\SU(2)_F$ flavor current.

\item[(vii)]  The final sum on the first line of \eqref{eq:LS-Index-IIB-3}, $-\sum_{k=1}^\infty t^{3(k+5)/2}\chi_{[1]}(y)\chi_{[k+1]}(x)$, corresponds to a tower of $L\bar{A}_2[\frac{3k+11}{4};\frac{1}{2},0;\frac{k+1}{2}]\otimes [\frac{k+1}{2}]$ multiplets.

\item[(viii)]  The second sum on the third line of \eqref{eq:LS-Index-IIB-3}, $-\sum_{k=0}^\infty t^{3(k+7)/2}\chi_{[k+1]}(x)$, corresponds to a tower of $L\bar{A}_2(\frac{3k+17}{4};0,0;\frac{k+1}{2})\otimes [\frac{k+1}{2}]$ multiplets. 
\end{enumerate}
The semi-short multiplets described in (i)-(viii) above correspond precisely to the KK towers of semi-short multiplets presented in Section~\ref{subsec:KKspecExFT}. This agreement between the QFT and supergravity calculations constitutes a non-trivial test of AdS/CFT. 

Interestingly we find that the two towers described in (iv) can actually be distinguished from each other in the supergravity computation in Section \ref{subsec:KKspecExFT}. The reason is that in supergravity there is an additional $\U(1)_R \times \U(1)_Y$ action under which the solution is invariant. The two towers are oppositely charged with respect to this $\U(1)_Y$, and can therefore be distinguished, as has been done in \eqref{Eq: spin12 semi-shorts}. Guided by the $\U(1)_Y$ bonus symmetry in the planar limit of $\mathcal{N}=4$ SYM, we expect that correlation functions of local operators in the planar LS SCFT enjoy a symmetry under the combined $\U(1)_R \times \U(1)_Y$ action discussed in Section \ref{sec:5dKK}. Since this is not a symmetry of the LS theory for general gauge groups there is no corresponding fugacity for it in the index calculation above. 

\section{Discussion}
\label{sec:discussion}

We have shown how to calculate the full KK spectrum of the AdS$_5$ PW solution of type IIB supergravity using recently developed ExFT techniques. We corroborated the results of this explicit calculation by two other direct methods. We computed the spectrum of all spin-2 KK modes by evaluating the eigenvalues for the regular solutions of the Laplace equation on the internal space. We also found the spectrum of short and semi-short multiplets in the LS theory by calculating the superconformal index. 

There are a number of interesting questions that arise in the context of our work that deserve further study. We discuss a few of them below.

\begin{itemize}

\item As discussed around \eqref{eq:LBmarginal}, the LS SCFT belongs to a conformal manifold of complex dimension 3. At a generic point on the conformal manifold the $\SU(2)_F$ flavor symmetry should be broken and there should be no continuous flavor symmetry. It will be very interesting to construct the family of AdS$_5$ supergravity solutions that are holographically dual to this conformal manifold and compute the corresponding spectrum of KK excitations. 

\item There is another 4d $\mathcal{N}=1$ SCFT which is closely related to the LS theory and should share many similarities in its operator spectrum. We refer to this theory as $\rm LS/\mathbb{Z}_2$ since it can be obtained by an RG flow from the 4d $\mathcal{N}=2$ two-node quiver SCFTs arising from a $\mathbb{Z}_2$ orbifold of $\mathcal{N}=4$ SYM. The holographic dual of this RG flow can be found by performing the corresponding $\mathbb{Z}_2$ orbifold of the PW type IIB solution \cite{Pilch:2000ej,Pilch:2000fu}. In the UV one simply finds the AdS$_5\times S^5/\mathbb{Z}_2$ solution, while in the IR one has a $\mathbb{Z}_2$ orbifold of the AdS$_5$ PW solution. The 4d $\mathcal{N}=1$ theory studied by Klebanov-Witten (KW) is another well-known RG flow that originates from the $\mathbb{Z}_2$ orbifold of $\mathcal{N}=4$ SYM and ends at an IR strongly coupled SCFT \cite{Klebanov:1998hh}. The gravitational dual of this SCFT is given by the well-known AdS$_5\times T^{1,1}$ solution of type IIB supergravity \cite{Romans:1984an}. The spectrum of KK excitations around this supergravity solution was calculated in detail in \cite{Ceresole:1999ht,Ceresole:1999zs}, see also \cite{Gubser:1998vd}, and, as shown in \cite{Nakayama:2005mf,Nakayama:2006ur,Gadde:2010en}, the part of the spectrum corresponding to protected SCFT multiplets agrees with the superconformal index. The $\rm LS/\mathbb{Z}_2$ and KW SCFTs belong to the same five-dimensional complex conformal manifold and are defined by specific superpotentials with enhanced flavor symmetry, see \cite{Corrado:2004bz,Benvenuti:2005wi}. This conformal manifold does not appear to have a weak coupling limit and therefore it is challenging to study how the operator spectrum of the SCFTs describing it depend on the exactly marginal couplings. It is possible that holography and KK spectroscopy may offer some insight into this question. There is a one-dimensional submanifold of the conformal manifold with an $\SU(2)$ flavor symmetry which has a known dual type IIB supergravity solution \cite{Halmagyi:2005pn}. Using the explicit KK spectrum of AdS$_5\times T^{11}$  \cite{Ceresole:1999ht,Ceresole:1999zs} and the KK spectrum of the PW solution computed here, together with the family of AdS$_5$ solutions in \cite{Halmagyi:2005pn} may allow for explicit calculations of the dimension of some of the SCFT operators as a function of an exactly marginal coupling. We plan to study this question further in \cite{BMRSvM}.

\item A notable feature of our results is that there are unprotected operators in the LS SCFT spectrum which are dual to KK supergravity modes and thus do not have large anomalous dimensions controlled by the 't Hooft coupling. Similar features were observed in \cite{Gubser:1998vd,Ceresole:1999ht,Ceresole:1999zs} for the spectrum of operators of the KW SCFT \cite{Klebanov:1998hh}. It is desirable to have an understanding of the QFT mechanism responsible for this feature of the operator spectrum. In addition to being of intrinsic QFT interest, this question is also important for understanding the emergence of large internal dimensions in AdS/CFT. A better understanding of this mechanism may elucidate the conditions under which scale separated AdS vacua can exist in string theory and supergravity, see \cite{Alday:2019qrf} for a recent discussion.

\item Some aspects of the KK spectrum calculation presented in Section~\ref{sec:5dKK} deserve better understanding. In particular the role of the $U(1)_Y$ ``bonus'' symmetry and its role in the ExFT formalism should be elucidated. In addition it will be nice to understand whether there is a deeper reason for the appearance of the simple generating functions we employed to organize the KK spectrum.

\item In Section~\ref{sec:index} we presented the calculation of what can be called the ``Hamiltonian index'', i.e. we have neglected the supersymmetric Casimir energy prefactor discussed in \cite{Assel:2015nca,Bobev:2015kza}. In addition, as in \cite{Gadde:2010en}, we have assumed that the fugacities appearing in the index are real when we studied the large $N$ limit. Both of these assumptions are not important when comparing the large $N$ limit of the index with the KK supergravity spectrum. However, as emphasized recently in a number of papers, including \cite{Hosseini:2017mds,Cabo-Bizet:2018ehj,Choi:2018hmj,Benini:2018ywd}, these modifications of the superconformal index could be crucial to understand the physics of supersymmetric black holes in AdS$_5$. It will be very interesting to understand the large $N$ limit of the LS SCFT in this context and to construct the dual AdS$_5$ black hole solutions of supergravity.

\item It is natural to wonder whether our KK spectroscopy results can be generalized to other known AdS$_5$ solutions of type IIB supergravity dual to 4d $\mathcal{N}=1$ SCFTs, like the AdS$_5\times Y^{p,q}$ solutions \cite{Gauntlett:2004yd} or the gravity dual of the $\beta$-deformation of $\mathcal{N}=4$ SYM
\cite{Lunin:2005jy}. Currently we do not know of a suitable consistent truncation of IIB supergravity to an extended supergravity theory in 5d for these classes of solutions to which we can apply the ExFT KK spectroscopy technique of \cite{Malek:2019eaz,Malek:2020yue}. Nevertheless, it is tempting to speculate that one can perhaps combine various techniques to attack this KK spectroscopy problem. For example one can utilize the results in \cite{Eager:2012hx}\footnote{The cyclic homology approach of \cite{Eager:2012hx} can also be used to compute the spectrum of protected KK modes of the Pilch-Warner solution discussed here, see \cite{Eager:2015hwa} for a brief discussion. We are grateful to Richard Eager for informing us about these results.} for the spectrum of protected operators in combination with a more explicit calculation of the eigenvalues of the Laplacian on $Y^{p,q}$ studied in \cite{Kihara:2005nt,Chen:2012wr} to compute the KK spectrum of spin-2 modes as we did in Section~\ref{sec:IIB}. Understanding these KK spectra in detail is an important open problem in the application of supergravity to the holographic study of 4d $\mathcal{N}=1$ SCFTs.

\end{itemize}

\noindent \textbf{ Acknowledgements }
\medskip

\noindent We are grateful to Fri\dh rik Freyr Gautason, Krzysztof Pilch, and Jan Rozendaal for useful discussions. NB and BR are supported in part by an Odysseus grant G0F9516N from the FWO. EM is supported by the Deutsche Forschungsgemeinschaft (DFG, German Research Foundation) via the Emmy Noether program ``Exploring the landscape of string theory flux vacua using exceptional field theory'' (project number 426510644). The work of JvM is supported by a doctoral fellowship from the Research Foundation - Flanders (FWO). NB, JvM, and BR are also supported by the KU Leuven C1 grant ZKD1118 C16/16/005. 

\begin{appendices}


\section{4d $\mathcal N=1$ superconformal multiplets}
\label{App: N=1 d=4 multiplets}
In this appendix we summarize some relevant facts about the representation theory and multiplet structure for theories with 4d $\mathcal{N}=1$ superconformal symmetry. We follow closely the presentation in \cite{Cordova:2016emh}.\footnote{See \cite{Dobrev:1985qv} for early original results on superconformal representation theory, which are also summarized in \cite{Dobrev:2004tk}.}

The 4d $\mathcal{N}=1$ superconformal algebra is $\SU(2,2|1)$. Its representations are labelled by the conformal dimension $\Delta$, the $\U(1)$ R-charge $r$, as well as the Lorentz spin quantum numbers $\left[ j_1, j_2 \right] \in \left[ \frac{\mathbb{N}_0}{2},\frac{\mathbb{N}_0}{2} \right]$.\footnote{In this appendix we do not include the representation of the operators under any flavor symmetry.} For a given state/operator, $\Phi$, in the 4d $\mathcal{N}=1$ SCFT we use the following notation to indicate this data
\begin{equation}
\Phi : \left[ j_1 , j_2 \right]_{\Delta}^{(r)}\,.
\end{equation}
The following representations of the 4d Lorentz group play an important role in our discussion of the KK spectroscopy\footnote{Note that we use Lorentz spin-labels instead of the $\SU(2)$ Dynkin labels in \cite{Cordova:2016emh}.}
\begin{equation}\label{eq:Lorentzspinfields}
\begin{aligned}
\text{Scalars}:& \hspace{0.4cm} \quad \phi : \left[0,0\right] \,,\\
\text{Spin-}1/2:&\hspace{0.4cm} \quad \lambda : \left[\tfrac12,0\right] \,, \hspace{0.05cm}\quad \text{and} \quad \bar\lambda : \left[0,\tfrac12\right] \,, \\
\text{Vectors}:& \hspace{0.15cm}\quad A_\mu : \left[\tfrac12,\tfrac12\right] \,,\\
\text{Two-forms}:& \quad B_{\mu\nu} : \left[1,0\right] \,, \hspace{0.2cm}\quad \text{and} \quad \bar B_{\mu\nu} : \left[0,1\right] \,, \\
\text{Spin-}3/2:& \hspace{0.2cm}\quad \psi_\mu : \left[1,\tfrac12\right] \,, \quad \text{and} \quad \bar \psi_{\mu} : \left[\tfrac12,1\right] \,,\\
\text{Spin-}2:& \hspace{0.15cm}\quad g_{\mu\nu} : \left[1,1\right] \,.
\end{aligned}
\end{equation}

The $\mathcal N=1$ superalgebra has four Poincar\'e and four conformal supercharges. The superconformal representations are then built from a superconformal primary operator annihilated by the special conformal generators and the conformal supercharges by acting with the four  Poincar\'e supercharges. The quantum numbers of the Poincar\'e supercharges are given explicitly by
\begin{equation}
\begin{aligned}
Q: \left[ \tfrac12,0 \right]^{(-1)}_{\frac12}\,, \qquad\qquad \bar Q: \left[ 0,\tfrac12 \right]^{(+1)}_{\frac12}\,.
\end{aligned}
\end{equation}

To denote the various superconformal multiplets we will use the notation
\begin{equation}
X\bar Y[\Delta;j_1 , j_2;r]\,.
\end{equation}
The letters $X\bar Y$ denote the type of shortening condition, if any, on the multiplet for which we use the results of \cite{Cordova:2016emh}.

There are the following type of 4d $\mathcal N=1$ superconformal multiplets:
\begin{itemize}
\item long multiplets, $L\bar L\left[ \Delta; j_1,j_2; r \right]$, for which, 
\begin{equation}\label{eq:LLdef}
\Delta > 2 + \text{max}\left( 2j_1-\tfrac32 r, 2j_2 +\tfrac32 r \right)\,.
\end{equation}
\item semi-short multiplets:
\begin{itemize}
\item[{\tiny $\blacksquare$}] $L\bar A_1\left[ \Delta; j_1,j_2 \geq \tfrac12 ;r \right]$ and $A_1 \bar L\left[ \Delta; j_1 \geq \tfrac12,j_2 ;r \right]$  with
\begin{equation}
\begin{aligned}
L\bar A_1:\qquad &\Delta = 2(1 + j_2) + \tfrac32 r \,, \quad \text{and} \quad r > \tfrac23 (j_1-j_2)\,,\\
A_1 \bar L:\qquad &\Delta = 2 (1 +  j_1) - \tfrac32 r \,, \quad \text{and} \quad r < \tfrac23 (j_1-j_2)\,.
\end{aligned}
\end{equation}
\item[{\tiny $\blacksquare$}] $L\bar A_2\left[ \Delta;j_1,0 ;r \right]$ and $A_2 \bar L\left[ \Delta;0,j_2 ;r \right]$  with
\begin{equation}
\begin{aligned}
L\bar A_2:\qquad &\Delta = 2  + \tfrac32 r \,, \quad \text{and} \quad r > \tfrac23 j_1\,, \\
A_2 \bar L: \qquad&\Delta = 2  - \tfrac32 r \,, \quad \text{and} \quad r <- \tfrac23 j_2\,.
\end{aligned}
\end{equation}
\item[{\tiny $\blacksquare$}] $L\bar B_1\left[ \Delta;j_1,0;r \right]$, and $B_1 \bar L\left[ \Delta;0,j_2;r \right]$  with
\begin{equation}
\begin{aligned}
&\Delta =  \tfrac32 r \,, \quad \text{and} \quad r > \tfrac23 (1+j_1)\,,\\
&\Delta =  - \tfrac32 r \,, \quad \text{and} \quad r < -\tfrac23 (1+j_2)\,.
\end{aligned}
\end{equation}
\end{itemize}
\item short multiplets:
\begin{itemize}
\item[{\tiny $\blacksquare$}] $A_1 \bar A_2\left[ \Delta; j_1\geq \frac12, 0 ; r\right]$ and $A_2 \bar A_1\left[ \Delta; 0, j_2 \geq \frac12; r\right]$  with
\begin{equation}
\begin{aligned}
A_1 \bar A_2: \qquad &\Delta = 2 +  j_1\,, \quad \text{and} \quad r =\tfrac23 j_1\,, \\
A_2 \bar A_1: \qquad &\Delta = 2 +  j_2\,, \quad \text{and} \quad r =-\tfrac23 j_2\,.
\end{aligned}
\end{equation}

\item[{\tiny $\blacksquare$}] $A_1 \bar A_1\left[ \Delta; j_1\geq \frac12, j_2 \geq \frac12 ; r \right]$ with
\begin{equation}
\Delta = 2 + j_1+j_2\,, \quad \text{and} \quad r = \tfrac23(j_1-j_2)\,.
\end{equation}
\item[{\tiny $\blacksquare$}] $A_2 \bar A_2\left[ \Delta;0, 0; r\right]$ with
\begin{equation}
\Delta = 2\,, \quad \text{and} \quad r =0\,.
\end{equation}
\item[{\tiny $\blacksquare$}] $A_1 \bar B_1\left[ \Delta; j_1\geq \frac12, 0;r\right]$ and $B_1 \bar A_1\left[ \Delta; 0, j_2 \geq \frac12 ;r\right]$  with
\begin{equation}
\begin{aligned}
A_1 \bar B_1: \qquad &\Delta = 1 +  j_1\,, \quad \text{and} \quad r =\tfrac23 (1+j_1)\,, \\
B_1 \bar A_1: \qquad &\Delta = 1 +  j_2\,, \quad \text{and} \quad r =-\tfrac23 (1+j_2)\,.
\end{aligned}
\end{equation}
\item[{\tiny $\blacksquare$}] $A_2 \bar B_1\left[ \Delta; 0, 0 ;r\right]$ and $B_1 \bar A_2\left[ \Delta; 0, 0 ;r\right]$  with
\begin{equation}
\begin{aligned}
A_2 \bar B_1: \qquad &\Delta = 1 \,, \quad \text{and} \quad r =\tfrac23 \,,\\
B_1 \bar A_2:\qquad &\Delta = 1 \,, \quad \text{and} \quad r =-\tfrac23 \,.
\end{aligned}
\end{equation}
\item[{\tiny $\blacksquare$}] $B_1 \bar B_1\left[ \Delta; 0, 0 ; r\right]$ which is simply the identity operator with 
\begin{equation}
\begin{aligned}
&\Delta = 0 \,, \quad \text{and} \quad r =0 \,.\\
\end{aligned}
\end{equation}
\end{itemize}
\end{itemize}

We have listed all short multiplets allowed by unitarity and 4d $\mathcal{N}=1$ superconformal symmetry. For the LS SCFT of interest in this work not all of these multiplets appear in the spectrum. In addition to the $B_1 \bar B_1\left[ 0; 0, 0 ; 0\right]$ identity operator we have only two other short multiplets.  We have an $A_1 \bar A_1\left[\Delta;\frac12,\frac12;r\right]$ multiplet which contains the energy-momentum tensor and the $\U(1)$ R-current operators and an $A_2 \bar A_2\left[ \Delta; 0,0 ;r\right]$ multiplet which transforms in the adjoint of the $\SU(2)_F$ flavor symmetry and contains the conserved flavor current. All the semi-short and the long multiplets listed above appear in the spectrum of the LS theory.

In the diagrams below we give the explicit field content of all the superconformal multiplets appearing in the spectrum of the LS theory. We use the same Lorentz spin notation as in \eqref{eq:Lorentzspinfields} and denote the R-charge of each field with a superscript.

\begin{equ}[H]
\begin{equation*}
\begin{aligned}
& \xymatrix @C=5.7pc @R=5.7pc @!0 @dr {
{A_\mu^{(0)}} \ar[r]|--{{~\bar Q~}} \ar[d]|--{~Q~} 
& {\bar\psi_{\mu}^{(+1)}}  \ar[d]|--{~Q~} \\
{\psi_{\mu}^{(-1)}  } \ar[r]|--{{~\bar Q~}}
& {g_{\mu\nu}^{(0)}} 
}
\end{aligned}
\quad
\begin{aligned}
& \xymatrix @C=5.7pc @R=5.7pc @!0 @dr {
{\phi^{(0)}} \ar[r]|--{{~\bar Q~}} \ar[d]|--{~Q~} 
& {\bar\lambda^{(+1)}}  \ar[d]|--{~Q~} \\
{\lambda^{(-1)}  } \ar[r]|--{{~\bar Q~}}
& {A_{\mu}^{(0)}} 
}
\end{aligned}
\end{equation*}
\caption{The $A_1\bar A_1  \left[ \Delta;\frac12,\frac12 ;0 \right]$ and $A_2 \bar A_2  \left[ \Delta;0 ,0 ;0 \right]$ multiplets.}\label{}
\end{equ}

\begin{equ}[H]
\begin{equation*}
\begin{aligned}
& \xymatrix @C=5.7pc @R=5.7pc @!0 @dr {
{\phi^{(r)}} \ar[d]|--{~Q~} \\
{\lambda^{(r-1)}} \ar[d]|--{~Q~} \\
{\phi^{(r-2)} }
}
\end{aligned}
\qquad
\begin{aligned}
& \xymatrix @C=5.7pc @R=5.7pc @!0 @dr {
{\phi^{(r)}} \ar[r]|--{{~\bar Q~}} 
& {\bar \lambda^{(r+1)}} \ar[r]|--{{~\bar Q~}} 
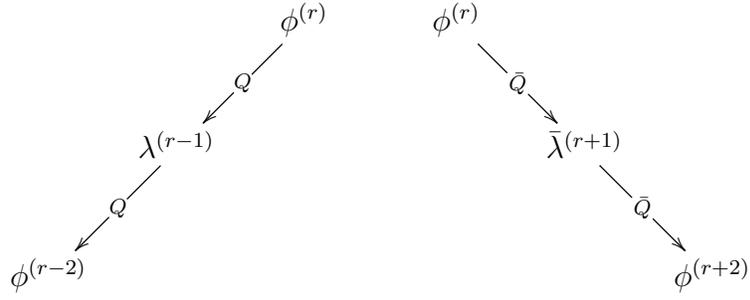
& {\phi^{(r+2)}}
}
\end{aligned}
\end{equation*}
\caption{The $L\bar B_1  \left[ \Delta;0,0 ;r \right]$ and $B_1 \bar L  \left[ \Delta;0 ,0 ;r \right]$ multiplets.}\label{}
\end{equ}

\begin{equ}[H]
\begin{equation*}
\begin{aligned}
& \xymatrix @C=5.7pc @R=5.7pc @!0 @dr {
{\lambda^{(r)}} \ar[d]|--{~Q~} \\
{B_{\mu\nu}^{(r-1)} + \phi^{(r-1)} } \ar[d]|--{~Q~} \\
{\lambda^{(r-2)} }
}
\end{aligned}
\qquad
\begin{aligned}
& \xymatrix @C=5.7pc @R=5.7pc @!0 @dr {
{\bar\lambda^{(r)}} \ar[r]|--{{~\bar Q~}} 
& { \bar B_{\mu\nu}^{(r+1)} + \phi^{(r+1)} } \ar[r]|--{{~\bar Q~}} 
& {\bar\lambda^{(r+2)}}
}
\end{aligned}
\end{equation*}
\caption{The $L\bar B_1  \left[ \Delta;\frac12,0 ;r \right]$ and $B_1 \bar L  \left[ \Delta;0 ,\frac12 ;r \right]$ multiplets.}\label{}
\end{equ}

\begin{equ}[H]
\begin{equation*}
\begin{aligned}
& \xymatrix @C=5.7pc @R=5.7pc @!0 @dr {
{\bar\lambda^{(r)}} \ar[r]|--{{~\bar Q~}} \ar[d]|--{~Q~} 
& {\bar B_{\mu\nu}^{(r+1)}}  \ar[d]|--{~Q~} \\
{A_\mu^{(r-1)}} \ar[r]|--{{~\bar Q~}} \ar[d]|--{~Q~} 
& {\bar\psi_{\mu}^{(r)}}\ar[d]|--{~Q~}\\
{\bar\lambda^{(r-2)} } \ar[r]|--{{~\bar Q~}}
& {\bar B_{\mu\nu}^{(r-1)}} 
}
\end{aligned}
\quad
\begin{aligned}
& \xymatrix @C=5.7pc @R=5.7pc @!0 @dr {
{\lambda^{(r)}} \ar[r]|--{{~\bar Q~}} \ar[d]|--{~Q~} 
& {A_\mu^{(r+1)}} \ar[r]|--{{~\bar Q~}} \ar[d]|--{~Q~} 
& {\lambda^{(r+2)}} \ar[d]|--{~Q~}\\
{B_{\mu\nu}^{(r-1)}} \ar[r]|--{{~\bar Q~}}
& {\psi_\mu^{(r)}}\ar[r]|--{{~\bar Q~}}
 & {B_{\mu\nu}^{(r+1)}} 
}
\end{aligned}
\end{equation*}
\caption{The $L\bar A_1  \left[ \Delta;0,\frac12 ;r \right]$ and $A_1 \bar L  \left[ \Delta;\frac12 ,0 ;r \right]$ multiplets.}\label{}
\end{equ}

\begin{equ}[H]
\begin{equation*}
\begin{aligned}
& \xymatrix @C=5.7pc @R=5.7pc @!0 @dr {
{A_\mu^{(r)}} \ar[r]|--{{~\bar Q~}} \ar[d]|--{~Q~} 
& {\bar \psi_{\mu}^{(r+1)}}  \ar[d]|--{~Q~} \\
{\psi_\mu^{(r-1)}+\bar \lambda^{(r-1)}} \ar[r]|--{{~\bar Q~}} \ar[d]|--{~Q~} 
& {g_{\mu\nu}^{(r)} + \bar B_{\mu\nu}^{(r)}}\ar[d]|--{~Q~}\\
{A_\mu^{(r-2)} } \ar[r]|--{{~\bar Q~}}
& {\bar \psi_{\mu}^{(r-1)}} 
}
\end{aligned}
\quad
\begin{aligned}
& \xymatrix @C=5.7pc @R=5.7pc @!0 @dr {
{A_\mu^{(r)}} \ar[r]|--{{~\bar Q~}} \ar[d]|--{~Q~} 
& { \bar \psi_\mu^{(r+1)} + \lambda^{(r+1)} } \ar[r]|--{{~\bar Q~}} \ar[d]|--{~Q~} 
& {A_\mu^{(r+2)}} \ar[d]|--{~Q~}\\
{\psi_{\mu}^{(r-1)}} \ar[r]|--{{~\bar Q~}}
& {g_{\mu\nu}^{(r)} + B_{\mu\nu}^{(r)} }\ar[r]|--{{~\bar Q~}}
 & {\psi_{\mu}^{(r+1)}} 
}
\end{aligned}
\end{equation*}
\caption{The $L\bar A_1  \left[ \Delta;\frac12,\frac12 ;r \right]$ and $A_1 \bar L  \left[ \Delta;\frac12 ,\frac12 ;r \right]$ multiplets.}\label{}
\end{equ}

\begin{equ}[H]
\begin{equation*}
\begin{aligned}
& \xymatrix @C=5.7pc @R=5.7pc @!0 @dr {
{\phi^{(r)}} \ar[r]|--{{~\bar Q~}} \ar[d]|--{~Q~} 
& {\bar \lambda_{\mu}^{(r+1)}}  \ar[d]|--{~Q~} \\
{ \lambda^{(r-1)}} \ar[r]|--{{~\bar Q~}} \ar[d]|--{~Q~} 
& {A_{\mu}^{(r)}}\ar[d]|--{~Q~}\\
{\phi^{(r-2)} } \ar[r]|--{{~\bar Q~}}
& {\bar \lambda^{(r-1)}} 
}
\end{aligned}
\quad
\begin{aligned}
& \xymatrix @C=5.7pc @R=5.7pc @!0 @dr {
{\phi^{(r)}} \ar[r]|--{{~\bar Q~}} \ar[d]|--{~Q~} 
& { \bar \lambda^{(r+1)} } \ar[r]|--{{~\bar Q~}} \ar[d]|--{~Q~} 
& {\phi^{(r+2)}} \ar[d]|--{~Q~}\\
{\lambda_{\mu}^{(r-1)}} \ar[r]|--{{~\bar Q~}}
& {A_{\mu}^{(r)} }\ar[r]|--{{~\bar Q~}}
 & {\lambda^{(r+1)}} 
}
\end{aligned}
\end{equation*}
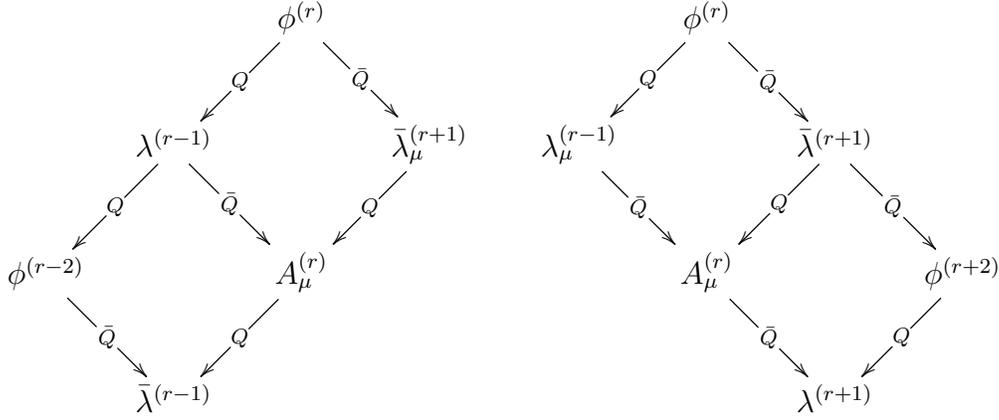
\caption{The $L\bar A_2  \left[ \Delta;0,0;r \right]$ and $A_2 \bar L  \left[ \Delta;0 ,0;r \right]$ multiplets.}\label{}
\end{equ}

\begin{equ}[H]
\begin{equation*}
\begin{aligned}
& \xymatrix @C=5.7pc @R=5.7pc @!0 @dr {
{\lambda^{(r)}} \ar[r]|--{{~\bar Q~}} \ar[d]|--{~Q~} 
& {A_{\nu}^{(r+1)}}  \ar[d]|--{~Q~} \\
{B_{\mu\nu}^{(r-1)} + \phi^{(r-1)} } \ar[r]|--{{~\bar Q~}} \ar[d]|--{~Q~} 
& {\psi_{\mu}^{(r)} + \bar\lambda^{(r)} }\ar[d]|--{~Q~}\\
{\lambda^{(r-2)} } \ar[r]|--{{~\bar Q~}}
& {A_{\mu}^{(r-1)}} 
}
\end{aligned}
\quad
\begin{aligned}
& \xymatrix @C=5.7pc @R=5.7pc @!0 @dr {
{\bar \lambda^{(r)}} \ar[r]|--{{~\bar Q~}} \ar[d]|--{~Q~} 
& {\bar B_{\mu\nu}^{(r+1)} + \phi^{(r+1)}} \ar[r]|--{{~\bar Q~}} \ar[d]|--{~Q~} 
& {\bar\lambda^{(r+2)}} \ar[d]|--{~Q~}\\
{A_{\mu}^{(r-1)}} \ar[r]|--{{~\bar Q~}}
& {\bar \psi_\mu^{(r)} +  \lambda^{(r)} }\ar[r]|--{{~\bar Q~}}
 & {A_{\mu}^{(r+1)}} 
}
\end{aligned}
\end{equation*}
\caption{The $L\bar A_2  \left[ \Delta;\frac12,0 ;r \right]$ and $A_2 \bar L  \left[ \Delta;0 ,\frac12 ;r \right]$ multiplets.}\label{}
\end{equ}

\begin{equ}[H]
\begin{equation*}
\begin{aligned}
& \xymatrix @C=5.7pc @R=5.7pc @!0 @dr {
{\lambda^{(r)}} \ar[r]|--{{~\bar Q~}} \ar[d]|--{~Q~} 
& {A_\mu^{(r+1)}} \ar[r]|--{{~\bar Q~}} \ar[d]|--{~Q~} 
& {\lambda^{(r+2)}} \ar[d]|--{~Q~}\\
{B_{\mu\nu}^{(r-1)} + \phi^{(r-1)}} \ar[r]|--{{~\bar Q~}} \ar[d]|--{~Q~} 
& {\lambda^{(r)} + \psi_\mu^{(r)} + \bar \lambda^{(r)}}\ar[r]|--{{~\bar Q~}}\ar[d]|--{~Q~}
 & {B_{\mu\nu}^{(r+1)} + \phi^{(r+1)}} \ar[d]|--{~Q~}\\
{\lambda^{(r-2)}} \ar[r]|--{{~\bar Q~}}
& {A_\mu^{(r-1)}} \ar[r]|--{{~\bar Q~}}  
& {\lambda^{(r)}} 
}
\end{aligned}
\quad
\begin{aligned}
& \xymatrix @C=5.7pc @R=5.7pc @!0 @dr {
{\bar\lambda^{(r)}} \ar[r]|--{{~\bar Q~}} \ar[d]|--{~Q~} 
& {\bar B_{\mu\nu}^{(r+1)} + \phi^{(r+1)}} \ar[r]|--{{~\bar Q~}} \ar[d]|--{~Q~} 
& {\bar\lambda^{(r+2)}} \ar[d]|--{~Q~}\\
{A_\mu^{(r-1)}} \ar[r]|--{{~\bar Q~}} \ar[d]|--{~Q~} 
& {\bar\lambda^{(r)} +\bar \psi_\mu^{(r)} +  \lambda^{(r)}}\ar[r]|--{{~\bar Q~}}\ar[d]|--{~Q~}
 & {A_\mu^{(r+1)}} \ar[d]|--{~Q~}\\
{\bar\lambda^{(r-2)}} \ar[r]|--{{~\bar Q~}}
& {\bar B_{\mu\nu}^{(r-1)} + \phi^{(r-1)}} \ar[r]|--{{~\bar Q~}}  
& {\lambda^{(r)}} 
}
\end{aligned}
\end{equation*}
\caption{The $L\bar L  \left[ \Delta;\frac12,0 ;r \right]$ and $L \bar L  \left[ \Delta;0 ,\frac12 ;r \right]$ multiplets. }\label{}
\end{equ}

\begin{equ}[H]
\begin{equation*}
\begin{aligned}
& \xymatrix @C=5.7pc @R=5.7pc @!0 @dr {
{\phi^{(r)}} \ar[r]|--{{~\bar Q~}} \ar[d]|--{~Q~} 
& {\bar \lambda^{(r+1)}} \ar[r]|--{{~\bar Q~}} \ar[d]|--{~Q~} 
& {\phi^{(r+2)}} \ar[d]|--{~Q~}\\
{\lambda^{(r-1)}} \ar[r]|--{{~\bar Q~}} \ar[d]|--{~Q~} 
& {A_{\mu}^{(r)}}\ar[r]|--{{~\bar Q~}}\ar[d]|--{~Q~}
 & {\lambda^{(r+1)}} \ar[d]|--{~Q~}\\
{\phi^{(r-2)}} \ar[r]|--{{~\bar Q~}}
& {\bar\lambda^{(r-1)}} \ar[r]|--{{~\bar Q~}}  
& {\phi^{(r)}} 
}
\end{aligned}
\quad
\begin{aligned}
& \xymatrix @C=5.7pc @R=5.7pc @!0 @dr {
{A_\mu^{(r)}} \ar[r]|--{{~\bar Q~}} \ar[d]|--{~Q~} 
& {\bar\psi_\mu^{(r+1)}+\lambda^{(r+1)}} \ar[r]|--{{~\bar Q~}} \ar[d]|--{~Q~} 
& {A_\mu^{(r+2)}} \ar[d]|--{~Q~}\\
{\psi_{\mu}^{(r-1)} + \bar\lambda^{(r-1)}} \ar[r]|--{{~\bar Q~}} \ar[d]|--{~Q~} 
& {\phi^{(r)} + B_{\mu\nu}^{(r)} +  \bar B_{\mu\nu}^{(r)} +  g_{\mu\nu}^{(r)}}\ar[r]|--{{~\bar Q~}}\ar[d]|--{~Q~}
 & {\psi_{\mu}^{(r+1)} + \bar\lambda^{(r+1)}} \ar[d]|--{~Q~}\\
{A^{(r-2)}} \ar[r]|--{{~\bar Q~}}
& {\bar\psi_\mu^{(r-1)}+\lambda^{(r-1)}} \ar[r]|--{{~\bar Q~}}  
& {A_\mu^{(r)}} 
}
\end{aligned}
\end{equation*}
\caption{The $L\bar L  \left[ \Delta;0,0 ;r \right]$ and $L \bar L  \left[ \Delta;\frac12 ,\frac12 ;r \right]$ multiplets.}\label{}
\end{equ}

\section{Explicit KK spectrum results}
\label{App: ExFT results}

Here we explicitly present part of the operator spectrum of the LS SCFT, derived using the ExFT techniques described in Section~\ref{subsec:ExFT}. All results below are presented in terms of 4d $\mathcal{N}=1$ superconformal multiplets using the notation in Appendix \ref{App: N=1 d=4 multiplets}. In the tables below we have also added the additional labels forf the $\SU(2)_F$ spin through $\otimes [k]$ as well as the $\U(1)_P \times \U(1)_Y$ charge via the superscript $s=(p+2y)$, i.e. we are using the notation
\begin{equation}
X\bar Y[\Delta;j_1,j_2;r]\otimes\left[k\right]^{(p+2y)}\,.
\end{equation}
%

In addition to the multiplets presented explicitly in the tables below one has to add their conjugates. For Tables~\ref{Tab: multiplets with vector primary} and \ref{Tab: multiplets with scalar primary} this means that multiplets of the form
\begin{equation}
Y \bar X [\Delta; \tfrac12,\tfrac12;-r] \otimes [k]^{(-[p+2y])}\,, \quad \text{and} \quad Y \bar X [\Delta; 0,0;-r] \otimes \left[k\right]^{(-[p+2y])}\,,
\end{equation}
have to be added whenever $r\neq 0$ or $s=p+2y\neq0$. For Table~\ref{Tab: multiplets with fermion primary} all multiplets have to be complemented with their conjugate multiplets of the form
\begin{equation}
Y \bar X [\Delta; 0,\tfrac12;-r] \otimes\left[k\right]^{(-[p+2y])}\,.
\end{equation}
It can be checked that the results in the tables below are indeed consistent with the analytic formulae presented in Section~\ref{subsec:KKspecExFT}. 

\begin{table}[H]
\makebox[\textwidth][c]{
	\renewcommand{\arraystretch}{1.4}
	\begin{tabular}{llll}
	 {$\scriptstyle\hspace{1.3cm}n=0$}																   	& {$\scriptstyle\hspace{1.3cm}n=1$} 	 																														& {$\scriptstyle\hspace{1.3cm}n=2$}  & {$\scriptstyle\hspace{1.3cm}n=3$}  \\ \hline\hline
		$\scriptstyle A_1 \bar A_1 [3;\tfrac12,\tfrac12;0] \otimes [0]^{(0)}$	 & $\scriptstyle L \bar A_1 [\tfrac92;\tfrac12;\tfrac12;1] \otimes [0]^{(0)}$   																&$\scriptstyle L \bar A_1 [6;\tfrac12;\tfrac12;2] \otimes [0]^{(0)}$   		& $\scriptstyle L \bar A_1 [\tfrac{15}{2};\tfrac12,\tfrac12;3] \otimes [0]^{(0)}$  \\
																											&   $\scriptstyle L \bar L [\tfrac{4+\sqrt{145}}{4};\tfrac12,\tfrac12;\tfrac12] \otimes [\tfrac12]^{(1/2)}$ 					 &$\scriptstyle L \bar L [\tfrac{2+\sqrt{61}}{2};\tfrac12;\tfrac12;1] \otimes [1]^{(1)}$ 	& $\scriptstyle L \bar L [\tfrac{23}{4};\tfrac12,\tfrac12;\tfrac32] \otimes [\tfrac32]^{(3/2)}$ 	\\
																											&				&  $\scriptstyle L \bar L [5;\tfrac12,\tfrac12;0] \otimes [1]^{(0)}$ 															& $\scriptstyle L \bar L [\tfrac{4+\sqrt{385}}{4};\tfrac12,\tfrac12;\tfrac12] \otimes [\tfrac32]^{(1/2)}$ 	\\
																											&				&$\scriptstyle L \bar L [\tfrac{4+\sqrt{337}}{4};\tfrac12,\tfrac12;\tfrac32] \otimes [\tfrac12]^{(1/2)}$	 	& $\scriptstyle L \bar L [1+\sqrt{31};\tfrac12,\tfrac12;2] \otimes [1]^{(1)}$   \\
																											&					&  $\scriptstyle L \bar L [\tfrac{4+\sqrt{313}}{4};\tfrac12,\tfrac12;\tfrac12] \otimes [\tfrac12]^{(-1/2)}$   & $\scriptstyle L \bar L [\tfrac{13}{2};\tfrac12,\tfrac12;1] \otimes [1]^{(0)}$  \\
																											&					& $\scriptstyle L \bar L [1+\sqrt{22};\tfrac12,\tfrac12;0] \otimes [0]^{(0)}$  											& $\scriptstyle L \bar L[1+2\sqrt{7};\tfrac12,\tfrac12;0] \otimes [1]^{(1)}$  \\
																											&																																										& 						& $\scriptstyle L \bar L [\tfrac{4+\sqrt{601}}{4};\tfrac12,\tfrac12;\tfrac52] \otimes [\tfrac12]^{(1/2)}$  \\
																											&																																										& 						& $\scriptstyle L \bar L [\tfrac{4+\sqrt{533}}{4};\tfrac12,\tfrac12;\tfrac32] \otimes [\tfrac12]^{(-1/2)}$  \\ 
																											&																																										&				  & $\scriptstyle L \bar L [\tfrac{27}{4};\tfrac12,\tfrac12;\tfrac12] \otimes [\tfrac12]^{(1/2)}$  \\
																											&																																										& 					 & $\scriptstyle L \bar L [\tfrac{2+\sqrt{145}}{2};\tfrac12,\tfrac12;1] \otimes [0]^{(0)}$  \\
	\end{tabular} 
}
\caption{Multiplets with vector primaries up to KK level $n=3$. }\label{Tab: multiplets with vector primary}
\end{table}
\begin{table}[H]
\makebox[\textwidth][c]{
	\renewcommand{\arraystretch}{1.2}
	\begin{tabular}{llll}
	 {$\scriptstyle \hspace{1.3cm}n=0$}																				& {$\scriptstyle \hspace{1.3cm}n=1$} 	 																   & {$\scriptstyle \hspace{1.3cm}n=2$} 																							 & {$\scriptstyle \hspace{1.3cm}n=3$}  \\ \hline\hline
		$\scriptstyle L \bar B_1 [\tfrac94;\tfrac12,0;\tfrac32] \otimes [\tfrac12]^{(3/2)}$				&   $\scriptstyle L \bar B_1 [3;\tfrac12,0;2] \otimes [1]^{(2)}$   									 & $\scriptstyle L \bar B_1 [\tfrac{15}{4};\tfrac12,0;\tfrac52] \otimes [\tfrac32]^{(5/2)}$  						& $\scriptstyle L \bar B_1 [\tfrac{9}{2};\tfrac12,0;3] \otimes [2]^{(3)}$ 	\\
		$\scriptstyle L \bar A_2 [\tfrac{11}{4};\tfrac12,0;\tfrac12] \otimes [\tfrac12]^{(1/2)}$	&   $\scriptstyle A_1 \bar L  [\tfrac{9}{2};\tfrac12,0;-1] \otimes [0]^{(\pm1)}$   			  &  $\scriptstyle A_1 \bar L [6;\tfrac12,0;-2] \otimes [0]^{(\pm1)}$										 						& $\scriptstyle A_1\bar L [\tfrac{15}{2};\tfrac12,0;-3] \otimes [0]^{(\pm1)}$ 	\\
		$\scriptstyle A_1 \bar L  [3;\tfrac12,0;0] \otimes [0]^{(1)}$	&	$\scriptstyle L \bar A_2 [\tfrac{7}{2};\tfrac12,0;1] \otimes [1]^{(1)}$ 		  			 &  $\scriptstyle L \bar A_2  [\tfrac{17}{4};\tfrac12,0;\tfrac32] \otimes [\tfrac32]^{(3/2)}$																 &	 $\scriptstyle L \bar A_2 [5;\tfrac12,0;2] \otimes [2]^{(2)}$   \\
																																	&  	$\scriptstyle L \bar L [1+\sqrt{7};\tfrac12,0;0] \otimes [1]^{(0)}$ 			 			   & 		 $\scriptstyle L \bar L [\tfrac{4+\sqrt{193}}{4};\tfrac12,0;\pm\tfrac12] \otimes [\tfrac32]^{(\pm1/2)}$ 					& $\scriptstyle L \bar L  [\tfrac{2+\sqrt{73}}{2};\tfrac12,0;\pm1] \otimes [2]^{(\pm1)}$  \\
																																	&  	$\scriptstyle L \bar L [\tfrac{15}{4};\tfrac12,0;\tfrac12] \otimes [\tfrac12]^{(3/2)}$ 	 &$\scriptstyle L \bar L [\tfrac92;\tfrac12,0;1] \otimes [1]^{(2)}$ 	& $\scriptstyle L \bar L[1+\sqrt{19};\tfrac12,0;0] \otimes [2]^{(0)}$  \\
																																	&	$\scriptstyle L \bar L [\tfrac{17}{4};\tfrac12,0;-\tfrac12] \otimes [\tfrac12]^{(1/2)}$  & 			$\scriptstyle L \bar L [\tfrac{2+\sqrt{73}}{2};\tfrac12,0;\pm1] \otimes [1]^{(0)}$													& $\scriptstyle L \bar L [\tfrac{21}{4};\tfrac12,0;\tfrac32] \otimes [\tfrac32]^{(5/2)}$  \\
																																	&	$\scriptstyle L \bar L [\tfrac{9}{2};\tfrac12,0;1] \otimes [0]^{(1)}$ 							& 			$\scriptstyle L \bar L [5;\tfrac12,0;0] \otimes [1]^{(1)}$					& $\scriptstyle L \bar L [\tfrac{4+\sqrt{433}}{4};\tfrac12,0;\pm\tfrac32] \otimes [\tfrac32]^{(\pm1/2)}$   \\
																																	&																															&  $\scriptstyle L \bar L [\tfrac{4+\sqrt{313}}{4};\tfrac12,0;\pm\tfrac32] \otimes [\tfrac12]^{(\pm3/2)}$  	&  $\scriptstyle L \bar L [\tfrac{23}{4};\tfrac12,0;\tfrac12] \otimes [\tfrac32]^{(3/2)}$   \\
																																	&																															&   $\scriptstyle L \bar L [\tfrac{4+\sqrt{337}}{4};\tfrac12,0;\pm\tfrac12] \otimes [\tfrac12]^{(\pm1/2)}$	 & $\scriptstyle L \bar L [\tfrac{4+\sqrt{409}}{4};\tfrac12,0;\pm\tfrac12] \otimes [\tfrac32]^{(\mp1/2)}$   \\
																																	&																															&  		 $\scriptstyle L \bar L [\tfrac{21}{4};\tfrac12,0;-\tfrac12] \otimes [\tfrac12]^{(3/2)}$  	 				&$\scriptstyle L \bar L [1+2\sqrt{7};\tfrac12,0;\pm2] \otimes [1]^{(\pm2)}$   \\
																																	&																															& 	 		$\scriptstyle L \bar L [\tfrac{23}{4};\tfrac12,0;-\tfrac32] \otimes [\tfrac12]^{(1/2)}$ 					&$\scriptstyle L \bar L [1+\sqrt{34};\tfrac12,0;\pm2] \otimes [1]^{(0)}$   \\
																																	&																															& 			$\scriptstyle L \bar L [6;\tfrac12,0;2] \otimes [0]^{(1)}$  														&  $\scriptstyle L \bar L [\tfrac{13}{2};\tfrac12,0;\pm1] \otimes [1]^{(\pm1)}$   \\
																																	&																															&  			$\scriptstyle L \bar L [1+\sqrt{22};\tfrac12,0;0] \otimes [0]^{(\pm1)}$  											  &  $\scriptstyle L \bar L [\tfrac{13}{2};\tfrac12,0;-1] \otimes [1]^{(1)}$   \\
																																	&																															&  																																						&$\scriptstyle L \bar L [6;\tfrac12,0;0] \otimes [1]^{(2)}$   \\
																																	&																															&																																						 & $\scriptstyle 2\times L \bar L [1+\sqrt{31};\tfrac12,0;0] \otimes [1]^{(0)}$ \\
																																	&																															&  																																						 &  $\scriptstyle L \bar L [\tfrac{4+\sqrt{577}}{4};\tfrac12,0;\pm\tfrac52] \otimes [\tfrac12]^{(\pm3/2)}$ \\
																																	&																															&  																																						 &  $\scriptstyle L \bar L [\tfrac{4+\sqrt{577}}{4};\tfrac12,0;\pm\tfrac32] \otimes [\tfrac12]^{(\pm1/2)}$    \\
																																	&																															& 																																						 &  $\scriptstyle L \bar L [\tfrac{4+\sqrt{505}}{4};\tfrac12,0;\pm\tfrac12] \otimes [\tfrac12]^{(\pm3/2)}$    \\
																																	&																															& 																																						&  $\scriptstyle L \bar L [\tfrac{4+\sqrt{553}}{4};\tfrac12,0;\pm\tfrac12] \otimes [\tfrac12]^{(\mp1/2)}$    \\
																																	&																															& 																																						 &  $\scriptstyle L \bar L [\tfrac{27}{4};\tfrac12,0;-\tfrac32] \otimes [\tfrac12]^{(3/2)}$   \\
																																	&																															& 																																						 &  $\scriptstyle L \bar L [\tfrac{29}{4};\tfrac12,0;-\tfrac52] \otimes [\tfrac12]^{(1/2)}$   \\
																																	&																															& 																																						 &  $\scriptstyle L \bar L [\tfrac{15}{2};\tfrac12,0;3] \otimes [0]^{(1)}$   \\
																																	&																															& 																																						 &  $\scriptstyle L \bar L [\tfrac{2+\sqrt{145}}{2};\tfrac12,0;\pm1] \otimes [0]^{(\pm1)}$   \\
																																	&																															& 																																						 &  $\scriptstyle L \bar L [\tfrac{2+\sqrt{145}}{2};\tfrac12,0;\pm1] \otimes [0]^{(\mp1)}$   \\
	\end{tabular} 
}
\caption{Multiplets with fermion primaries up to KK level $n=3$.}\label{Tab: multiplets with fermion primary}
\end{table}
\begin{table}[H]
\makebox[\textwidth][c]{
	\renewcommand{\arraystretch}{1.1}
	\begin{tabular}{llll}
	 {$\scriptstyle \hspace{1.3cm}n=0$}												& {$\scriptstyle \hspace{1.3cm}n=1$} 	 																						& {$\scriptstyle \hspace{1.3cm}n=2$}  & {$\scriptstyle \hspace{1.3cm}n=3$}  \\ \hline\hline
		$\scriptstyle L \bar B_1 [\tfrac32;0,0;1] \otimes [1]^{(1)}$	&   $\scriptstyle L \bar B_1 [\tfrac94;0,0;\tfrac32] \otimes [\tfrac32]^{(3/2)}$   							  & $\scriptstyle L \bar B_1 [3;0,0;2] \otimes [2]^{(2)}$  																	& $\scriptstyle L \bar B_1 [\tfrac{15}{4};0,0;\tfrac52] \otimes [\tfrac52]^{(5/2)}$ 	\\
		$\scriptstyle L \bar B_1 [3;0,0;2] \otimes [0]^{(2)}$				&   $\scriptstyle L \bar B_1 [\tfrac{15}{4};0,0;\tfrac52] \otimes [\tfrac12]^{(5/2)}$   			 	 &  $\scriptstyle L \bar B_1 [\tfrac92;0,0;3] \otimes [1]^{(3)}$														& $\scriptstyle L \bar B_1 [\tfrac{21}{4};0,0;\tfrac72] \otimes [\tfrac32]^{(7/2)}$ 	\\
		$\scriptstyle L\bar L [1+\sqrt{7};0,0;0] \otimes [0]^{(0)}$  	&	$\scriptstyle L \bar A_2 [\tfrac{11}{4};0,0;\tfrac12] \otimes [\tfrac32]^{(1/2)}$					&$\scriptstyle L \bar A_2 [\tfrac72;0,0;1] \otimes [2]^{(1)}$														   & $\scriptstyle L \bar A_2 [\tfrac{17}{4};0,0;\tfrac32] \otimes [\tfrac52]^{(3/2)}$   \\
																											&  $\scriptstyle L \bar A_2 [\tfrac{17}{4};0,0;\tfrac32] \otimes [\tfrac12]^{(3/2)}$					  & $\scriptstyle L \bar A_2  [5;0,0;2] \otimes [1]^{(2)}$ 																	& $\scriptstyle L \bar A_2 [\tfrac{23}{4};0,0;\tfrac52] \otimes [\tfrac32]^{(5/2)}$  \\
																											&$\scriptstyle L \bar L [1+\sqrt{7};0,0;0] \otimes [1]^{(1)}$   														& $\scriptstyle L \bar L [1+\sqrt{7};0,0;0] \otimes [2]^{(0)}$ 														 & $\scriptstyle L \bar L[\tfrac{4+\sqrt{193}}{4};0,0;\tfrac12] \otimes [\tfrac52]^{(1/2)}$  \\
																											&	 $\scriptstyle L \bar L [\tfrac{4+\sqrt{193}}{4};0,0;\tfrac12] \otimes [\tfrac12]^{(1/2)}$ 		 & $\scriptstyle L \bar L [\tfrac{17}{4};0,0;\tfrac12] \otimes [\tfrac32]^{(3/2)}$ 							& $\scriptstyle L \bar L [\tfrac{2+\sqrt{61}}{2};0,0;1] \otimes [2]^{(2)}$  \\
																											&	$\scriptstyle L \bar L [\tfrac{2+\sqrt{37}}{2};0,0;1] \otimes [0]^{(2)}$ 								   & $\scriptstyle L \bar L [\tfrac{4+\sqrt{217}}{4};0,0;\tfrac12] \otimes [\tfrac32]^{(-1/2)}$		 & $\scriptstyle L \bar L [\tfrac{2+\sqrt{85}}{2};0,0;1] \otimes [2]^{(0)}$   \\
																											&	$\scriptstyle L \bar L [\tfrac{2+\sqrt{61}}{2};0,0;1] \otimes [0]^{(0)}$ 								   &  $\scriptstyle 2 \times L \bar L [\tfrac{2+\sqrt{73}}{2};0,0;1] \otimes [1]^{(1)}$  				 &  $\scriptstyle L \bar L [1+\sqrt{19};0,0;0] \otimes [2]^{(1)}$   \\
																											&																																							  & $\scriptstyle L \bar L [\tfrac{2+\sqrt{73}}{2};0,0;1] \otimes [1]^{(-1)}$								 & $\scriptstyle 2\times  L \bar L [\tfrac{4+\sqrt{409}}{4};0,0;\tfrac32] \otimes [\tfrac32]^{(3/2)}$   \\
																											&																																							& 	$\scriptstyle L \bar L [1+\sqrt{19};0,0;0] \otimes [1]^{(0)}$	&$\scriptstyle L \bar L [\tfrac{4+\sqrt{457}}{4};0,0;\tfrac32] \otimes [\tfrac32]^{(-1/2)}$   \\
																											&																																							&   $\scriptstyle L \bar L [\tfrac{4+\sqrt{241}}{4};0,0;\tfrac32] \otimes [\tfrac12]^{(5/2)}$	&$\scriptstyle 2\times  L \bar L [\tfrac{4+\sqrt{433}}{4};0,0;\tfrac12] \otimes [\tfrac32]^{(1/2)}$   \\
																											&																																							&  $\scriptstyle L \bar L [\tfrac{4+\sqrt{385}}{4};0,0;\tfrac32] \otimes [\tfrac12]^{(1/2)}$  	&  $\scriptstyle L \bar L [\tfrac{4+\sqrt{385}}{4};0,0;\tfrac12] \otimes [\tfrac32]^{(-3/2)}$   \\
																											&																																							& 	  $\scriptstyle L \bar L [\tfrac{4+\sqrt{313}}{4};0,0;\tfrac12] \otimes [\tfrac12]^{(3/2)}$ 					&  $\scriptstyle L \bar L [1+\sqrt{22};0,0;2] \otimes [1]^{(3)}$   \\
																											&																																							& 	  $\scriptstyle L \bar L [\tfrac{23}{4};0,0;\tfrac12] \otimes [\tfrac12]^{(-1/2)}$  											  &$\scriptstyle 2\times  L \bar L [1+\sqrt{34};0,0;2] \otimes [1]^{(1)}$   \\
																											&																																							&  $\scriptstyle L \bar L [1+\sqrt{22};0,0;2] \otimes [0]^{(2)}$   													  & $\scriptstyle L \bar L [1+\sqrt{34};0,0;2] \otimes [1]^{(-1)}$ \\
																											&																																							&	  $\scriptstyle L \bar L [1+2\sqrt{7};0,0;2] \otimes [0]^{(0)}$  															  &  $\scriptstyle L \bar L [\tfrac{2+\sqrt{109}}{2};0,0;1] \otimes [1]^{(2)}$ \\
																											&																																							&  					  $\scriptstyle L \bar L [1+\sqrt{19};0,0;0] \otimes [0]^{(2)}$  					&  $\scriptstyle L \bar L [\tfrac{2+\sqrt{133}}{2};0,0;1] \otimes [1]^{(0)}$   \\
																											&																																							& 		$\scriptstyle 2\times L \bar L [6;0,0;0] \otimes [0]^{(0)}$  																																								 &  $\scriptstyle 2 \times L \bar L [1+\sqrt{31};0,0;0] \otimes [1]^{(1)}$   \\
																											&																																							& 																																						 &  $\scriptstyle L \bar L [\tfrac{4+\sqrt{505}}{4};0,0;\tfrac52] \otimes [\tfrac12]^{(5/2)}$   \\
																											&																																							& 																																						 &  $\scriptstyle L \bar L [\tfrac{4+\sqrt{649}}{4};0,0;\tfrac52] \otimes [\tfrac12]^{(1/2)}$   \\
																											&																																							& 																																						 &  $\scriptstyle L \bar L [\tfrac{4+\sqrt{553}}{4};0,0;\tfrac32] \otimes [\tfrac12]^{(3/2)}$   \\
																											&																																							& 																																						 &  $\scriptstyle L \bar L [\tfrac{4+\sqrt{601}}{4};0,0;\tfrac32] \otimes [\tfrac12]^{(-1/2)}$   \\
																											&																																							& 																																						 &  $\scriptstyle L \bar L [\tfrac{4+\sqrt{433}}{4};0,0;\tfrac12] \otimes [\tfrac12]^{(5/2)}$   \\
																											&																																							& 																																						 &  $\scriptstyle 2\times  L \bar L [\tfrac{4+\sqrt{577}}{4};0,0;\tfrac12] \otimes [\tfrac12]^{(1/2)}$   \\
																											&																																							& 																																						 &  $\scriptstyle L \bar L [\tfrac{27}{4};0,0;\tfrac12] \otimes [\tfrac12]^{(-3/2)}$   \\
																											&																																							& 																																						 &  $\scriptstyle L \bar L [\tfrac{2+\sqrt{157}}{2};0,0;3] \otimes [0]^{(2)}$   \\
																											&																																							& 																																						 &  $\scriptstyle L \bar L [\tfrac{2+\sqrt{181}}{2};0,0;3] \otimes [0]^{(0)}$   \\
																											&																																							& 																																						 &  $\scriptstyle L \bar L [\tfrac{2+\sqrt{133}}{2};0,0;1] \otimes [0]^{(\pm2)}$   \\
																											&																																							& 																																						 & $\scriptstyle 2 \times  L \bar L [\tfrac{2+\sqrt{157}}{2};0,0;1] \otimes [0]^{(0)}$   \\
	\end{tabular} 
}
\caption{Multiplets with scalar primaries up to KK level $n=3$.}\label{Tab: multiplets with scalar primary}
\end{table}
%

\end{appendices}

\bibliography{LSKKspec}
\bibliographystyle{JHEP}

\end{document}